\documentclass[twocolumn]{svjour3}          
\smartqed  
\usepackage{amsfonts}
\usepackage{amsmath}
\usepackage{amssymb}
\usepackage{algorithm}
\usepackage[noend]{algpseudocode}
\usepackage{bm} 
\usepackage{dsfont} 
\usepackage{booktabs}
\makeatletter
\let\cl@chapter\relax
\makeatother
\usepackage[capitalise]{cleveref} 
\usepackage{xspace} 
\usepackage{multirow} 
\usepackage{multicol} 
\usepackage[labelformat=simple]{subcaption}
\usepackage{adjustbox}
\usepackage{graphicx} 
\usepackage[numbers,sort&compress]{natbib}
\usepackage[table]{xcolor} 
\usepackage{url}

\newcommand{\spara}[1]{\smallskip\noindent{\bf #1}}

\newcommand{\mpara}[1]{\medskip\noindent{\bf #1}}

\newenvironment{squishlist}
{\begin{list}{$\bullet$}
 {\setlength{\itemsep}{0pt}
     \setlength{\parsep}{3pt}
     \setlength{\topsep}{3pt}
     \setlength{\partopsep}{0pt}
     \setlength{\leftmargin}{1.5em}
     \setlength{\labelwidth}{1em}
     \setlength{\labelsep}{0.5em} } }
{\end{list}}

\newcommand{\papertitle}{\our: Lossless Hypergraph Summarization via Co-Clustering\xspace}

\newcommand{\leman}{\textsc{CoClusLSH}\xspace}
\newcommand{\our}{\textsc{HyDRA}\xspace} 
\newcommand{\cc}{\textsc{PB-}$\tau$\textsc{CC}\xspace}
\newcommand{\tcc}{\textsc{PB-}$\tau$\textsc{TCC}\xspace}

\newcommand{\suchthat}{\ensuremath{\mathrel{:}}\xspace} 

\newcommand{\degrf}{\ensuremath{\mathsf{d}}\xspace} 
\newcommand{\degr}[1]{\ensuremath{\degrf(#1)}\xspace} 
\newcommand{\hdegrf}{\ensuremath{\mathsf{hd}}\xspace} 
\newcommand{\hdegr}[1]{\ensuremath{\hdegrf(#1)}\xspace} 

\newcommand{\nodes}{\ensuremath{\mathcal{V}}\xspace} 
\newcommand{\edges}{\ensuremath{\mathcal{E}}\xspace} 
\newcommand{\numnodes}{\ensuremath{\left|\mathcal{V}\right|}\xspace} 
\newcommand{\supernodes}{\ensuremath{\mathcal{P}}\xspace} 
\newcommand{\superedges}{\ensuremath{\mathcal{R}}\xspace} 
\newcommand{\hyperg}{\ensuremath{\mathcal{H}}\xspace} %
\newcommand{\summary}{\ensuremath{\mathcal{S}}\xspace} %

\newcommand{\lgrayo}{\cellcolor{gray!15} 1} 
\newcommand{\lgrayn}[1]{\cellcolor{gray!15} #1} 
\newcommand{\mgrayo}{\cellcolor{gray!45} 1} 
\newcommand{\dgrayo}{\cellcolor{gray!95} 1} 
\newcommand{\dgrayn}[1]{\cellcolor{gray!95} #1} 
\newcommand{\bgrayo}{\cellcolor{black!80} \textcolor{white}{1}} 
\newcommand{\erro}{\cellcolor{red!20} 1} 
\newcommand{\errn}[1]{\cellcolor{red!20} #1} 
\newcommand{\errz}{\cellcolor{red!60}} 

\begin{document}

\title{\papertitle}


\author{Giulia Preti
    \and
    Aris Anagnostopoulos
    \and
    Francesco Bonchi
}


\institute{Giulia Preti \at
           Intesa Sanpaolo Innovation Center, Turin, Italy \\
           \email{giulia.preti@intesasanpaolo.com}
           \and
           Aris Anagnostopoulos \at
           Sapienza University of Rome, Italy \\
           \email{aris@diag.uniroma1.it}
           \and
           Francesco Bonchi \at
           Intesa Sanpaolo AI Research, Turin, Italy \\
           \email{francesco.bonchi@intesasanpaolo.com}
}

\date{Received: date / Accepted: date}

\maketitle

\begin{abstract}
Hypergraphs are a powerful representation for higher-order interactions
but their  scale  and complexity pose significant data management and
analysis challenges. While summarization techniques are widely used to
distill simple graphs, lossless summarization for hypergraphs remains
unexplored. We introduce \our, the first formal framework for lossless
summarization of weighted hypergraphs. In our framework, a summary is a
new weighted hypergraph composed of \emph{supernodes} (groups of nodes)
and \emph{superhyperedges} (groups of hyperedges), paired with a
correction table for exact reconstruction. By establishing a
conceptual link to co-clustering, we design 
an efficient, parameter-free greedy algorithm that iteratively merges
node and hyperedge clusters to minimize a novel storage-aware cost
function. 
\our employs an incremental update strategy to prevent the costly
recomputation of the correction table at each step. Extensive
experiments demonstrate that \our achieves a substantial reduction in
storage cost (80-93\% in some settings, depending on the hypergraph characteristics). Because the resulting summaries are themselves
hypergraphs, they can be queried directly,
providing fast and accurate approximate answers for various connectivity
and centrality queries, and accelerating downstream tasks
such as influence maximization.
\end{abstract}

\section{Introduction}\label{sec:intro}
Biological, physical, and social systems often exhibit higher-order interactions, beyond the binary relations that are typically modeled by simple graphs.
In such scenarios, hypergraphs provide a powerful representation of such multi-way interactions.
Examples include
cellular processes~\cite{ritz2014pathway}, protein interaction
networks~\cite{feng2021hypergraph}, brain
activity~\cite{petri2014homological}, co-authorship
networks~\cite{patania2017shape}, contact
networks~\cite{billings2019simplex2vec,preti2024higher}, and
communication and social networks~\cite{yang2019revisiting}.
The growing availability of higher-order data has urged the development
of a diverse array of methodologies aimed at tackling the challenges
posed by higher-order interactions~\cite{lee2025survey}.
These include higher-order centrality scores~\cite{benson2019three},
clustering and
classification~\cite{benson2018simplicial,zhou2006learning}, hypergraph
neural networks~\cite{feng2019hypergraph}, higher-order link
prediction~\cite{zhang2018beyond}, pattern mining and motif
analysis~\cite{lotito2022higher,preti2022fresco},
recommendations~\cite{zheng2018novel,ji2020dual}, core/truss
decomposition~\cite{sun2020fully,preti2021strud}, and distance
oracles~\cite{preti2024hyper}.

However, the scale and complexity of hypergraphs pose significant
challenges in terms of data management, analysis, and interpretation.
Similarly to the case of graphs, \emph{summarization} can help distill
the essential information from hypergraphs.

\spara{Graph summarization.} The earliest graph summarization proposals
appeared concurrently at SIGMOD 2008
\cite{navlakha2008graph,tian2008efficient}. Navlakha et
al.~\cite{navlakha2008graph}
proposed a graph summary comprising (1) a coarse-grained graph
with supernodes (sets of nodes) and superedges representing the
relationships between these sets, and (2) a correction table
storing the information needed to losslessly  recreate the original
graph from the summary.
\emph{We adopt this formulation in our work.}

Concurrently, Tian et al.~\cite{tian2008efficient} introduced a method
to create a summary graph by grouping nodes that are homogeneous in
attributes and relationships, allowing the users to control the summary
resolution.
Later, 
LeFevre and Terzi~\cite{lefevre2010grass} introduced a \emph{lossy}
summarization
approach based on a probabilistic query semantics, trading off summary
size and accuracy.
Building on \cite{lefevre2010grass}, Riondato et
al.~\cite{riondato2017graph} proposed an algorithm with \emph{formal
quality guarantees}, ensuring that summaries preserve structural
properties within provable approximation bounds. We refer to the
survey by Liu et al.~\cite{liu2018graph} and the tutorials by Khan et
al.~\cite{khan2017summarizing} and Koutra et
al.~\cite{koutra2018summarizing} for comprehensive overviews of this
extensive literature.

%
%
%
%
%
%

Despite extensive work on \emph{graph} summarization,
\emph{to the best of our knowledge, we are the first to tackle the
problem of building a meaningful lossless summary of a given
hypergraph}.

\spara{Paper contributions and roadmap.}
This work introduces \our, the first formal framework for
\emph{lossless summarization of weighted hypergraphs}. Inspired by the
original graph summarization definition by Nav\-lakha et
al.~\cite{navlakha2008graph}, we formalize the problem of lossless
hypergraph summarization. In our setting, a lossless summary of an input
weighted hypergraph is itself a weighted hypergraph, such that its
nodes, called \emph{supernodes}, are groups of nodes of the input; its
hyperedges, called \emph{superhyperedges}, are sets of supernodes
representing groups of hyperedges of the input. Similarly to graph
summarization~\cite{navlakha2008graph}, a \emph{correction table} stores
the minimal information required for exact reconstruction of the
original input. To guide the search toward efficient and effective
summaries, we define an optimization problem that minimizes a
\emph{storage-aware cost function}, explicitly accounting for the space
required by superhyperedges, their weights, and the corrections.

As a first step towards developing a solution,
we establish a conceptual link between hypergraph summarization and
\emph{co-clustering}~\cite{hartigan1972direct}, showing that both aim to
identify coherent groups of rows and columns in a matrix representation
of the data.
This connection provides a theoretical foundation that allows us to
rethink hypergraph summarization as a co-clustering task, exploiting
existing algorithmic ideas.
As in our problem the optimal number of node and hyperedge clusters is
not known a priori, we develop a greedy algorithm, \our, which
iteratively merges node and hyperedge clusters that reduce the storage
cost, leveraging an incremental update strategy that avoids recomputing
the correction table from scratch at each iteration. \our is inspired by
the information-theoretic co-clustering framework of Gao and
Akoglu~\cite{gao2014fast}, which exhibits the desirable features of
optimizing an explicit objective function and not requiring the number
of clusters in input.

We evaluate \our against several baselines, including its original formulation, and parameter-free and spectral co-clustering algorithms.
Our experimental results demonstrate that \our produces summaries that
more effectively preserve the structure and connectivity patterns of the
input hypergraphs while achieving greater storage efficiency than
competing methods.
Because our summaries are themselves hypergraphs, they can be directly
queried—either exactly, by reconstructing the original hypergraph, or
approximately, by executing queries directly on the summary. This
property enables substantial computational savings while retaining
accuracy.
We show this through several connectivity and centrality queries,
including degree, closeness centrality, distance, and reachability.
In all cases, queries performed on the summaries yield results that
closely approximate those obtained on the input hypergraphs, while
requiring, in most cases, only a fraction of the time.
Finally, we present a case study on influence maximization~\cite{KKT},
demonstrating the practical utility of our summaries.

\smallskip
\noindent Our contributions can be summarized as follows:
\begin{squishlist}
    \item We formalize the problem of lossless hypergraph summarization,
    introducing the notions of supernodes, superhyperedges, and
    corrections that enable exact reconstruction (\Cref{sec:pb}).
    \item We design a storage-aware cost function that quantifies the
    space required for a summary, accounting for superhyperedges and
    correction tuples (\Cref{sec:pb}).
    \item We identify a natural connection between hypergraph
    summarization and co-clustering, providing a new perspective that
    allows us to leverage existing techniques in a higher-order setting
    (\Cref{sec:algo}).
    \item We develop \our, an efficient greedy algorithm that integrates
    our objective function with incremental update and post-processing
    strategies to produce valid summaries (\Cref{sec:algo}).
    \item We conduct extensive experiments on real-world hypergraphs,
    comparing our approach against several baselines. Results show that
    our method produces summaries that achieve a good tradeoff between
    storage reduction and query accuracy (\Cref{sec:exp}).
\end{squishlist}

\section{Related Work} \label{app:rw}

\spara{Graph summarization.}
Graph summarization has received significant attention thanks to applications in visualization, query processing, and large-scale mining. We refer to the survey by Liu et al.~\cite{liu2018graph} and the tutorials by Khan et al.~\cite{khan2017summarizing} and Koutra et al.~\cite{koutra2018summarizing} for comprehensive overviews; below, we cover the key works.

The earliest graph summarization proposals appear\-ed concurrently at SIGMOD 2008 \cite{navlakha2008graph,tian2008efficient}. Navlakha et al.~\cite{navlakha2008graph}
propose a graph summary comprising (\emph{i}) a coarse-grained graph with supernodes representing sets of nodes and superedges representing the relationships between these sets, and (\emph{ii}) a correction table storing the information needed to perfectly recreate the original graph from the coarse-grained graph, enabling lossless summarization.
This formulation forms the foundation of our work.
Concurrently, Tian et al.~\cite{tian2008efficient} introduce a method to create a summary graph by grouping nodes that are homogeneous in attributes and relationships, allowing the users to control the summary resolution.

LeFevre and Terzi~\cite{lefevre2010grass} introduce a \emph{lossy} summari\-zation
approach based on a probabilistic query semantics, trading off summary size and accuracy.
They show\-ed that common queries can be answered efficiently and accurately from these summaries.
Feng et al.~\cite{feng2012summarization} introduce a method for summarizing \emph{bipartite graphs} by partitioning nodes into groups that maximize information-theoretic quality measures. The summary is lossy, i.e., the original graph cannot be exactly recovered.
Riondato et al.~\cite{riondato2017graph} propose an algorithm with \emph{formal quality guarantees}, ensuring that summaries preserve structural properties within provable approximation bounds.
Ke et al.~\cite{ke2022multi} study \emph{multi-relation graph summarization}, capturing heterogeneous edge types by jointly considering multiple relationships between nodes.
Lee et al.~\cite{lee2022slugger} develop \emph{SLUGGER}, a lossless hierarchical summarization approach that recursively merges nodes while enabling exact reconstruction of the input.
Merchant et al.~\cite{merchant2023graph} design a \emph{spectral algorithm} that summarizes graphs via node grouping, achieving state-of-\-the-\-art structure preservation.
Chu et al.~\cite{chu2024graph} propose a scalable algorithm that balances \emph{compactness and efficiency}, generating summaries suitable for various downstream tasks.
Ali et al.~\cite{ali2024ssag} study \emph{attributed graph summarization}, introducing methods that jointly sparsify edges and compress attribute information.

While these methods effectively capture pairwise connectivity, they cannot distinguish between a set of nodes forming a single higher-order relation and the same set connected through multiple pairwise edges.
This limitation prevents direct application of graph summarization techniques to hypergraphs, where higher-order semantics must be explicitly preserved.

\spara{Summarization in databases.}  Database summarization aims to succinctly represent a collection of transactions, or a binary matrix. As a hypergraph incidence matrix is a binary matrix, this line of work is related.
Gionis et al.~\cite{gionis2004geometric} formulate summarization as finding \emph{tiles} -- rectangular regions of a binary matrix with associated probabilities -- that concisely explain observed data under a \emph{minimum description length (MDL)}-based scoring function. A tile can be viewed as a frequent itemset, but nodes can belong to multiple tiles, so no node partitioning is produced.
Mampaey and Vreeken \cite{mampaey2010summarising} propose a greedy bottom-up algorithm based on the MDL principle, merging items into clusters that minimize the encoded length of the database.
The output does not yield a node partitioning.
Xiang et al.~\cite{xiang2011summarizing} introduce \emph{overlapped hyperrectangles} to cover the transaction space exactly, enabling applications to bipartite graph summarization, but again without enforcing node partitioning.
Mampaey et al.~\cite{mampaey2011tell,mampaey2012summarizing} extend this approach by mining sets of \emph{informative itemsets} that best describe the database, producing succinct but overlapping representations.


\spara{Hypergraphs compression and sparsification.}
Although hypergraph summarization, compression, and sparsification all reduce the complexity of large hypergraphs, they target distinct goals.
Summarization focuses on creating a condensed representation that preserves the essential structural and semantic properties of the original hypergraph, and that can be queried and analyzed directly.
In contrast, compression~\cite{chierichetti2009compressing,boldi2011layered} aims to reduce the data size by encoding the hypergraph in a more compact form, often at the cost of losing some detail or requiring decompression for analysis.
Sparsification~\cite{bansal2019new}, on the other hand, removes less significant nodes or hyperedges to produce a sparser version of the input hypergraph.
Therefore, unlike compression, a hypergraph summary remains directly queryable; unlike sparsification, it retains all nodes and the structural coherence of the original hypergraph.

\spara{Co-Clustering.}
Hypergraph summarization is closely related to co-clustering, which jointly partitions the rows and columns of a matrix. This connection arises because a hypergraph can be equivalently represented via its incidence matrix.
Wang et al.~\cite{10.1145/3681793} and Battaglia et al.~\cite{10.1145/3698875} provide comprehensive surveys on co-clustering, covering classical algorithms, recent advances, and open problems.

Puolamäki et al.~\cite{puolamaki2008approximation} study biclustering and provide theoretical approximation guarantees for grouping rows and columns simultaneously.
Anagnostopoulos et al.~\cite{anagnostopoulos2012constant} introduce a constant-factor approximation algorithm requiring the number of row and column clusters, which is often difficult to set a priori.
Ramírez and Tepper~\cite{ramirez2013bi} propose an MDL-based matrix factorization framework for biclustering, providing a probabilistic interpretation of co-clusters. This method, however, does not produce a partitioning of the nodes.
Gao and Akoglu~\cite{gao2014fast} design a fast agglomerative algorithm that uses information-theoretic principles to build co-clusters in a scalable manner.
Hess et al.~\cite{hess2021broccoli} propose a biclustering algorithm based on proximal stochastic gradient descent that explicitly allows overlapping clusters and outliers. Unlike our formulation, which requires partitions of nodes and hyperedges, their approach does not enforce exclusivity, i.e., both node and hyperedge clusters may overlap.
Battaglia et al.~\cite{battaglia2024fast} propose a parameterless prototype-based co-clustering algorithm that achieves state-of-the-art performance without requiring hyperparameter tuning.

In this body of work, the approach of Battaglia et al.~\cite{battaglia2024fast} represents our main competitor.
The hypergraph summarization algorithm that we devise in this work (Section \ref{sec:algo}) follows the same general logic as the algorithm by Gao and Akoglu~\cite{gao2014fast}, but we introduce a different objective function and additional optimizations to improve scalability.
Among existing co-clust\-ering pa\-ra\-digms, their framework provides the most suitable foundation for our setting: it optimizes a well-defined objective function, and it does not require the number of clusters as input. These characteristics align closely with our requirements for lossless hypergraph summarization, where the number of clusters is unknown a priori.

\section{Problem Statement}\label{sec:pb}
We consider an undirected weighted \emph{hypergraph} $\hyperg \doteq
(\nodes,\edges,\omega)$ where $\nodes = \{v_1,\dots, v_n\}$ is a set of
nodes, $\edges = \{h_1,\dots, h_m\}$ is a set of \emph{hyperedges}, with each $h_i \subseteq \nodes$,
and $\omega: \edges \rightarrow \mathbb{N}^+$ is a weighting
function assigning a positive integer weight to each hyperedge.
For a node $v \in \nodes$, we define (1) the \emph{neighbor set}
$\Gamma_v \doteq \{u \in \nodes\setminus\{v\} \suchthat \exists h \in
\edges \suchthat u,v \in h \}$, which is the set of nodes that are in a
hyperedge with $v$; (2) the \emph{degree} $\degr{v} \doteq
\left|\Gamma_v\right|$, which is the number of its neighbors; and the
\emph{hyper-degree} $\hdegr{v} \doteq \sum_{\substack{h \in \edges}}
\omega(h) \mathds{1}_{v \in h}$, which is the weighted sum of the
hyperedges containing $v$.
%
%
A hypergraph $\hyperg$ can be represented via an \emph{incidence
matrix} $A \in \mathbb{N}^{n \times m}$, where $a_{ij} = \omega(j)$ if $v_i \in h_j$, and $0$ otherwise.

We are interested in \emph{lossless} summaries, which are compact
representations of $\hyperg$ that enable the exact reconstruction of $\hyperg$ through a set of \emph{corrections} applied to the summary.

\begin{figure}[!th]
    \centering
    \begin{subfigure}[c]{.49\linewidth}
    \centering
    \resizebox{\linewidth}{!}{
        \begin{tabular}{rc @{\hskip .25in} rc}
            & \textcolor{gray}{$\nodes$} & & \textcolor{gray}{$\nodes$} \\
            & \rule{.4in}{0.4pt} & & \rule{.4in}{0.4pt} \\
            \textcolor{gray}{0} & $\{0\}$ & \textcolor{gray}{6} & $\{6\}$ \\
            \textcolor{gray}{1} & $\{1\}$ & \textcolor{gray}{7} & $\{7\}$ \\
            \textcolor{gray}{2} & $\{2\}$ & \textcolor{gray}{8} & $\{8\}$ \\
            \textcolor{gray}{3} & $\{3\}$ & \textcolor{gray}{9} & $\{9\}$ \\
            \textcolor{gray}{4} & $\{4\}$ & \textcolor{gray}{10} & $\{10\}$ \\
            \textcolor{gray}{5} & $\{5\}$ & \textcolor{gray}{11} & $\{11\}$ \\
            \bottomrule
        \end{tabular}
    }
    \resizebox{\linewidth}{!}{
        \begin{tabular}{rcr}
            & \textcolor{gray}{$\edges$} & \textcolor{gray}{$\omega$} \\
            & \rule{1in}{0.4pt} & \rule{.1in}{0.4pt} \\
            \textcolor{gray}{0} & $\{1,2,3,4,5,6\}$ & 1 \\
            \textcolor{gray}{1} & $\{1,2,4,5\}$ & 1 \\
            \textcolor{gray}{2} & $\{2,3,4,5\}$ & 1 \\
            \textcolor{gray}{3} & $\{1,2,3,4,5\}$ & 1 \\
            \textcolor{gray}{4} & $\{1,4,5,7,0\}$ & 5 \\
            \textcolor{gray}{5} & $\{1,2,3,6,7,0,10,11\}$ & 1 \\
            \textcolor{gray}{6} & $\{1,2,3,6,7,10,11\}$ & 1 \\
            \textcolor{gray}{7} & $\{2,3,6,7,10,11\}$ & 1 \\
            \textcolor{gray}{8} & $\{1,2,6,7,8,10,11\}$ & 1 \\
            \textcolor{gray}{9} & $\{2,3,6,7,9,10,11\}$ & 3 \\
            \textcolor{gray}{10} & $\{1,2,6,7,10,11\}$ & 1 \\
            \textcolor{gray}{11} & $\{6,7,8,9,0\}$ & 3 \\
            \textcolor{gray}{12} & $\{4,6,7,8,9\}$ & 1 \\
            \textcolor{gray}{13} & $\{4,5,6,7,9,0\}$ & 1 \\
            \textcolor{gray}{14} & $\{6,8,9,0\}$ & 1 \\
            \textcolor{gray}{15} & $\{8,9,0\}$ & 1 \\
            \textcolor{gray}{16} & $\{2,8,9,0\}$ & 1 \\
            \textcolor{gray}{17} & $\{3,7,8,9\}$ & 1 \\
            \textcolor{gray}{18} & $\{6,7,0\}$ & 1 \\
        \end{tabular}
    }
    \end{subfigure}
    \begin{subfigure}[c]{.49\linewidth}
    \centering
    \resizebox{.59\linewidth}{!}{
              \begin{small}
        \begin{tabular}{rc}
            & \textcolor{gray}{$\supernodes$} \\
            \cline{2-2}
            \textcolor{gray}{0} & $\{1,2,3\}$\\
            \textcolor{gray}{1} & $\{4,5\}$\\
            \textcolor{gray}{2} & $\{6,7\}$\\
            \textcolor{gray}{3} & $\{8,9,0\}$\\
            \textcolor{gray}{4} & $\{10, 11\}$\\
            \bottomrule
        \end{tabular}
         \end{small}
    }
    \resizebox{.59\linewidth}{!}{
        \begin{tabular}{rcr}
            & \textcolor{gray}{$\superedges$} & \textcolor{gray}{$\omega_\summary$} \\
            \cline{2-3}
            \textcolor{gray}{0} & $\{0,1\}$ & 5 \\
            \textcolor{gray}{1} & $\{0,2,4\}$ & 6 \\
            \textcolor{gray}{2} & $\{2,3\}$ & 8 \\
            \textcolor{gray}{2} & $\{3\}$ & 2 \\
            \bottomrule
        \end{tabular}
    }
    \resizebox{.59\linewidth}{!}{
          \begin{small}
        \begin{tabular}{rc}
             & \textcolor{gray}{$\mathcal{C}$} \\
            \cline{2-2}
            \multirow{4}{*}{0}
              & $6$, \\
              & , $3$ \\
              & , $1$ \\
              & $7 \: 0$, $2 \: 3$, $5$ \\
            \cline{2-2}
            \multirow{5}{*}{1}
              & $0$ \\
              & , $1$ \\
              & $8$, $3$ \\
              & $9$, $1$, $3$ \\
              & , $3$ \\
            \cline{2-2}
            \multirow{5}{*}{2}
              & $4$, $0$ \\
              & $4 \: 5$, $8$ \\
              & , $7$ \\
              & $3$, $6 \: 0$ \\
              & , $8 \: 9$ \\
            \cline{2-2}
            \multirow{1}{*}{3}
              & $2$, \\
        \end{tabular}
              \end{small}
    }
    \end{subfigure}
    \caption{A hypergraph $\hyperg$ (left) and its summary $\summary$ (right).
    The hypergraph $\hyperg \doteq (\nodes, \edges)$ consists of a set of nodes and a set of weighted hyperedges.
    The summary $\summary \doteq (\supernodes, \superedges, \omega_\summary)$ consists of a set of supernodes
    $\supernodes = \boldsymbol{\{}P_0=\{1,2,3\}$ $,$ $P_1=\{4,5\}$~$,$ $P_2=\{6,7\}$ $,$ $P_3=\{8,9,0\}$ $,$ $P_4=\{10,11\}\boldsymbol{\}}$; a set of superhyperedges
	  $\superedges$ $=$ $\boldsymbol{\{}r_0=\{0,1\}$ $,$ $r_1=\{0,2,4\}$ $,$ $r_2=\{2,3\}$ $,$ $r_3=\{3\} \boldsymbol{\}}$ with weights
	  $\omega_\summary(r_0)=5$, $\omega_\summary(r_1)=6$, $\omega_\summary(r_2)=8$, $\omega_\summary(r_3)=2$; and a set of corrections
	  $\mathcal{C}=\boldsymbol{\{}C_0, C_1, C_2, C_3\boldsymbol{\}}$ with
	  $C_2=\{(\{4\}, \{0\}),$ $(\{4, 5\},$ $\{8\}), (\emptyset, \{7\}), (\{3\},\{0,6\}), (\emptyset,\{8,9\})\}$. The other $C_i$s are defined similarly.
    The cost of storing $\hyperg$ is $19 + 99 + 0 = 118$ and the cost of
    storing $\summary$ is $4 + 8 + 28 = 40$.
	}
    \label{fig:example}
\end{figure}

\begin{definition}[Lossless Summary]
A \emph{lossless summary} $\summary$ of a weighted hypergraph $\hyperg
\doteq (\nodes,\edges,\omega)$ consists of a weighted hypergraph
$(\supernodes, \superedges, \omega_\summary)$ together with a set of corrections
$\mathcal{C}$, where:
\begin{squishlist}
    \item $\supernodes = \left\{P_1, \cdots P_k\right\}$ is a
    partitioning of $\nodes$ into \emph{supernodes}. That is, $P_i
    \subseteq \nodes$, $\bigcup_{i=1}^k P_i = \nodes$, and $P_i \cap P_j
    = \emptyset$ for each $i \neq j$.
    \item $\superedges$ is a set of hyperedges over $\supernodes$,
    called \emph{superhyperedges}, with each $h \in\edges$ represented
    by \textbf{exactly one} superhyperedge $r \in \superedges$. Each
    superhyperedge $r$ has a positive integer weight
    $\omega_\summary(r)$.
    \item $\mathcal{C} = \{C_r : r \in \superedges\}$ contains sets of
    \emph{corrections} $c = (c^+, c^-, w)$, where (1) $c^+$ are
    nodes in the original hyperedge but not in the superhyperedge,
    (2) $c^-$ are nodes in the superhyperedge but not in the
    original hyperedge, and (3) $w$ is the weight of the
    original hyperedge (stored only if $w > 1$). Each $C_r$ allows the
    reconstruction of the hyperedges represented by $r$.
\end{squishlist}
\end{definition}
\noindent \Cref{fig:example} shows a hypergraph (left) and its lossless
summary (right), and its incidence matrix is illustrated in \Cref{fig:adj}.

The reconstruction of the original hypergraph $\hyperg$ from $\summary$ 
is obtained by setting $\edges = \emptyset$, then for each $r \in \superedges$ and each $c \in C_r$: 
(1) Create a hyperedge $h = \bigcup_{P \in r} P$, which is the union of the supernodes in $r$;
(2) Add the nodes in $c^+$ to $h$: $h \gets h \cup c^+$;
(3) Remove the nodes in $c^-$ from $h$: $h \gets h \setminus c^-$;
(4) Assign weight $w$ to $h$: $\omega(h) = w$. Note that, to save space, $c$ stores the weight $w$ of the original hyperedge only if $w > 1$. Thus, if $w$ is not stored, we set $\omega(h) = 1$;
(5) Add $h$ to $\edges$.
Finally, if $\omega_\summary(r) > |C_r|$, add the hyperedge $h = \bigcup_{P \in r} P$ to $\edges$ with weight $\omega(h) = \omega_\summary(r) - |C_r|$.

\begin{figure*}[!th]
    \centering
    \resizebox{.75\linewidth}{!}{
    \begin{tabular}{c|ccccc !{\vrule width 1.5pt} cccccc !{\vrule width 1.5pt} cccccc !{\vrule width 1.5pt} cc}
        & \textcolor{gray}{0} & \textcolor{gray}{1} & \textcolor{gray}{2} & \textcolor{gray}{3} & \textcolor{gray}{4} & \textcolor{gray}{5} & \textcolor{gray}{6} & \textcolor{gray}{7} & \textcolor{gray}{8} & \textcolor{gray}{9} & \textcolor{gray}{10} & \textcolor{gray}{11} & \textcolor{gray}{12} & \textcolor{gray}{13} & \textcolor{gray}{14} & \textcolor{gray}{17} & \textcolor{gray}{18} & \textcolor{gray}{15} & \textcolor{gray}{16} \\
        \cline{1-20}
        \textcolor{gray}{8} & & & & & & & & & \erro & & & \cellcolor{gray!45} 3 & \mgrayo & \errz & \mgrayo & \mgrayo & \errz & \bgrayo & \bgrayo \\
        \textcolor{gray}{9} & & & & & & & & & & \errn{3} & & \cellcolor{gray!45} 3 & \mgrayo & \mgrayo & \mgrayo  & \mgrayo & \errz & \bgrayo & \bgrayo \\
        \textcolor{gray}{0} & & & & & \errn{5} & \erro & & & & & & \cellcolor{gray!45} 3 & \errz & \mgrayo & \mgrayo & \errz & \mgrayo & \bgrayo & \bgrayo \\
        \cmidrule[1.5pt]{1-20}
        \textcolor{gray}{6} & \erro & & & & & \dgrayo & \dgrayo & \dgrayo & \dgrayo & \dgrayn{3} & \dgrayo & \cellcolor{gray!45} 3 & \mgrayo & \mgrayo & \mgrayo & \errz  & \mgrayo & & \\
        \textcolor{gray}{7} & & & & & \errn{5} & \dgrayo & \dgrayo & \dgrayo & \dgrayo & \dgrayn{3} & \dgrayo & \cellcolor{gray!45} 3 & \mgrayo & \mgrayo & \errz & \mgrayo & \mgrayo & & \\
        \cmidrule[1.5pt]{1-20}
        \textcolor{gray}{10} & & & & & & \dgrayo & \dgrayo & \dgrayo & \dgrayo & \dgrayn{3} & \dgrayo & & & & & & \\
        \textcolor{gray}{11} & & & & & & \dgrayo & \dgrayo & \dgrayo & \dgrayo & \dgrayn{3} & \dgrayo & & & & & & \\
        \cmidrule[1.5pt]{1-20}
        \textcolor{gray}{1} & \lgrayo & \lgrayo & \errz  & \lgrayo & \lgrayn{5} & \dgrayo & \dgrayo & \errz  & \dgrayo & \errz & \dgrayo & & & & & & & \\
        \textcolor{gray}{2} & \lgrayo & \lgrayo & \lgrayo & \lgrayo & \errz & \dgrayo & \dgrayo & \dgrayo & \dgrayo & \dgrayn{3} & \dgrayo & & & & & & & & \erro \\
        \textcolor{gray}{3} & \lgrayo & \errz  & \lgrayo & \lgrayo & \errz & \dgrayo & \dgrayo & \dgrayo & \errz & \dgrayn{3} & \errz & & & & & \erro & & &  \\
        \cmidrule[1.5pt]{1-20}
        \textcolor{gray}{4} & \lgrayo & \lgrayo & \lgrayo & \lgrayo & \lgrayn{5} & & & & & & & & \erro & \erro & & & \\
        \textcolor{gray}{5} & \lgrayo & \lgrayo & \lgrayo & \lgrayo & \lgrayn{5} & & & & & & & & & \erro & & & \\
    \end{tabular}
    }
    \caption{Incidence matrix of $\hyperg$ with rows (nodes) and columns (hyperedges) reordered to visualize the superhyperedges of $\summary$. Each superhyperedge is shown in a different shade of gray. Red cells mark corrections in $\mathcal{C}$: dark red for corrections of type $c^-$ and light red for type $c^+$.}
    \label{fig:adj}
\end{figure*}

The \emph{lossless hypergraph summarization problem} seeks a minimum-cost summary
of the input hypergraph:\footnote{We do not consider the cost of storing the nodes
because it is constant and equivalent for any summary/ hypergraph.}

\begin{problem}[\textsc{Lossless-HSum}]\label{pb:hsum}
Given a weighted hypergraph $H \doteq (\nodes, \edges, \omega)$, find a lossless summary $\summary \doteq (\supernodes, \superedges, \omega_\summary)$ with corrections $\mathcal{C}$ that minimizes the storage cost:
\begin{align}\label{eq:mincost}
\text{cost}(\summary) = \overset{\textcolor{red}{\emph{(a)}}}{|\superedges|} + \overset{\textcolor{red}{\emph{(b)}}}{\sum\limits_{r \in \superedges} |r|} + \overset{\textcolor{red}{\emph{(c)}}}{\sum\limits_{C \in \mathcal{C}}\sum\limits_{c \in C} \text{size}(c)}\,,
\end{align}
where $\text{size}(c) = |c^+| + |c^-| + \mathds{1}_{c \text{ stores } w}$.
\end{problem}

We require that each $h \in \edges$ is represented by exactly one $r \in \superedges$ to guarantee that $\superedges$ represents a \emph{partitioning} of the hyperedges, hence ensuring that the summary is well-defined and non-redundant.
On the other hand, the three terms in \Cref{eq:mincost} reflect different components of the storage cost: \textcolor{red}{(a)} is the cost of storing the weights of
    the superhyperedges (preventing the creation of multiple identical superhyperedges);
    \textcolor{red}{(b)} is the cost of storing the
    superhyperedges as lists of supernodes: this term encourages the reuse of supernodes and avoids trivial one-node supernodes that fail to exploit the hypergraph structure; \textcolor{red}{(c)} captures the deviations between the superhyperedges and the original hyperedges, hence quantifying the total information needed to achieve exact reconstruction. This term prevents degenerate solutions such as merging all nodes into one supernode (which would minimize \textcolor{red}{(a)} and \textcolor{red}{(b)} but make corrections nearly as large as the original hypergraph).

Two extreme scenarios illustrate the effect of these terms:

\spara{Maximal compression.}
If the summary has a single superhyperedge and a single supernode, the storage cost is
$1 + 1 + \left(\sum_{h \in \edges} (\numnodes - |h|) + \sum_{h \in \edges} \omega(h) - 1\right)$,
where $\sum_{e \in \edges} (\numnodes - |e|)$ is the cost of storing the sets $c^+$ and $c^-$, and $\sum_{e \in \edges} \omega(e) - 1$ is the cost of storing the hyperedge weights greater than $1$.
This summary has minimal superhyperedge cost but large correction costs.

\spara{No compression.}
If each hyperedge and node form their own superhyperedge and supernode (i.e., the summary equals the original hypergraph), the storage cost is $|\edges| + \sum_{h \in \edges} |h|$, yielding zero correction cost but maximal superhyperedge cost.

Together, the three terms \textcolor{red}{(a)--(c)} balance the
tradeoff between reusing structures (supernodes and superhyperedges)
and storing fine-grained differences (corrections), encouraging
summaries that are compact yet lossless.

\section{\our: A Solution to \textsc{Lossless-HSum}}\label{sec:algo}

The \textsc{Lossless-HSum} problem defined in \Cref{sec:pb} is closely
connected to existing work on \emph{co-clustering}.
Indeed, representing a weighted hypergraph via its incidence matrix $A
\in \mathbb{N}^{n \times m}$---where rows correspond to nodes and
columns to hyperedges---makes our goal of grouping nodes and hyperedges
equivalent to finding coherent row--column partitions of $A$.
Hence, constructing a hypergraph summary is similar to a form of
\emph{co-clustering}, where row clusters correspond to \emph{supernodes}
and column clusters to \emph{superhyperedges}.

Existing co-clustering methods differ mainly in their optimization
principles.
Information-theoretic methods minimize the loss of mutual information
between rows and columns \cite{dhillon2003information};
graph-theoretic methods perform bipartite spectral partitioning by
minimizing normalized cuts \cite{dhillon2001co}; and
matrix-factorization methods approximate the input matrix using low-rank
decompositions such as non-negative matrix tri-factorization (NMTF) \cite{del2015non}.
However, these techniques are unsuitable for our setting for two main reasons.
First, our summary must be a \emph{hypergraph}: each superhyperedge
must be a subset of supernodes, and supernodes are allowed to appear in
more than one superhyperedge.
Existing co-clustering methods produce row and column cluster assignments, but they do not define the explicit \emph{supernode--super}\-\emph{hyperedge} incidence required to reconstruct a valid hypergraph.
Second, our objective is to \emph{minimize storage cost}, which depends
on the number and size of superhyperedges, and on the correction table
needed for lossless reconstruction.
Therefore, even though one could, in principle, derive a summary from the output
of these methods, the result would not necessarily be compact, as their
objectives are not designed with hypergraph structure or storage efficiency in mind.

Despite these differences, the methodological parallels between
co-clustering and hypergraph summarization allow us to adapt and extend
existing algorithms to develop a new method specifically designed for
\emph{lossless and storage-efficient hypergraph summarization}.

Among existing co-clustering paradigms, we build upon the
information-theoretic agglomerative co-clust\-ering algorithm of Gao and
Akoglu~\cite{gao2014fast}.
This method, called \leman, minimizes an MDL-based objective through
greedy merges of similar row and column clusters, identified via
locality-sensitive hashing (LSH).
\leman offers two key advantages that align with our requirements:
(1) it optimizes an explicit cost function, which makes it
easier to integrate our own storage-based objective; and (2)
it does not require the number of clusters as input.

\mpara{From MDL to storage-aware cost.}
Gao and Akoglu's objective function quantifies the number of bits
required to encode the row and column cluster assignments (\emph{model
cost}) and the blocks of the adjacency matrix $A$ (\emph{data
cost}):
\[
\begin{split} 
L(A;R,C) &=\overset{\textcolor{red}{(\nu)}}{\log^* n} + \log^* m +
\overset{\textcolor{red}{(\nu)}}{\log^* k} +
\overset{\textcolor{red}{(\alpha)}}{\log^* \ell} \\
&+ \overset{\textcolor{red}{(\nu)}}{\sum_i r_i \log_2 \frac{n}{r_i}} + \sum_j c_j \log_2 \frac{m}{c_j} \\
&+ \sum_{i,j} \bigl( \overset{\textcolor{red}{(\beta)}}{\log_2(r_i c_j + 1)} + \overset{\textcolor{red}{(\gamma)}}{E(B_{ij})} \bigr),\,
\end{split}
\]
where $\log^*$ is the code length for integers, $n$ is the number of
rows, $m$ is the number of columns, $k$ is the number of row clusters,
$\ell$ is the number of column clusters, $r_i$ is the size of
of row cluster $i$, $c_j$ is the size of column cluster $j$, $\log_2(r_i
c_j + 1)$ is the number of bits required to encode the number of
$1$s in the block $B_{ij}$, and $E(B_{ij})$ is the number of bits
required to encode the block $B_{ij}$ itself (see~\cite{gao2014fast} for
its exact definition).

Although this objective function shares structural analogies with our
storage-cost objective (\Cref{eq:mincost}), the two serve different
goals. Their correspondence can be summarized as follows:
\begin{squishlist}
    \item Terms \textcolor{red}{($\nu$)} encode the cost of storing the
    row clusters. In our case, this corresponds to storing the
    number of nodes in each supernode. Given that the total cost equals
    to the sum over all supernodes, which in our case is constant across
    all possible summaries---equal to the total number of nodes---we omit
    it from our objective function.
    \item Term \textcolor{red}{($\alpha$)} accounts for the number of
    column clusters $\ell$, similar to our Term \textcolor{red}{(a)},
    which counts the number of superhyperedges.
    \item Term \textcolor{red}{($\beta$)} measures the cost of encoding
    the number of ones in each block, similar to our Term
    \textcolor{red}{(b)}, which captures the supernode--superhyperedge
    memberships.
    \item Term \textcolor{red}{($\gamma$)} encodes the content of each
    block, similar to our Term~\textcolor{red}{(c)}, which stores the
    corrections for ones outside and zeros inside each superhyperedge.
\end{squishlist}
Some terms, however, have no equivalent in our formulation: we do not
explicitly encode column-cluster assignments because hyperedges can be
reconstructed from the summary when needed, without preserving their
original membership structure.

\spara{Summarization-aware objective function.}
Our hypergraph summarization objective (Problem \ref{pb:hsum},
Eq.~\eqref{eq:mincost}) seeks to minimize the storage cost of the
summary, including superhyperedges, their incident supernodes, and the
correction table—including the weights of the original hyperedges, when
needed.
To adapt the MDL principle to this setting, we replace the
information-theoretic encoding length $L(A;R,C)$ with a relaxation of
our summarization cost, where we omit the term accounting for the
storage of hyperedge weights in the correction table.
This relaxation preserves the same tradeoff between model compactness
and reconstruction fidelity, while allowing the incremental computation
of cost updates via matrix operations.
This significantly reduces the per-merge cost, a reduction that is
crucial given the high number of merge operations required in
large-scale hypergraphs.

\subsection{Algorithm Overview}

Our algorithm, \our (\textbf{Hy}pergraph \textbf{D}ecomposition and
\textbf{R}eduction \textbf{A}lgorithm), illustrated in \Cref{alg:algorithm}, iteratively merges node and hyperedge clusters. 
 
We use a
\emph{density matrix} $D$ to cluster the rows and columns of $A$, which
will eventually define the supernodes and the superhyperedges of the hypergraph.
Initially, each node and hyperedge forms its own cluster, and
matrix $D$ is initialized as $A$.
Each row of $D$ represents a node cluster, each column a hyperedge
cluster, and each entry $D[i,j]$ records the fraction of non-zero
entries in the submatrix of $A$ induced by the node cluster $P_i \in
\mathcal{P}$ and hyperedge cluster $Q_j \in \mathcal{Q}$, that is,
\[
D[i,j] = \frac{\sum_{v \in P_i} \sum_{h \in Q_j} A[v,h]}{|P_i| \cdot |Q_j|}\,.
\]
Matrix $D$ serves two purposes: it accelerates merge-cost updates
and eventually will determine supernode inclusion in
superhyperedges. Specifically, supernode $P_i$ is included in the
superhyperedge generated from $Q_j$ if and only if $D[i,j] \geq 0.5$.
$D$ is updated after each merge to reflect the new inter-cluster densities.

\begin{algorithm}[!ht]
    \caption{\our}
    \label{alg:algorithm}
    \begin{algorithmic}[1]
    \Require Hypergraph Incidence Matrix $A$
    \Require LSH parameters $r, b$, Cost threshold $\tau$
    \Ensure A heuristic solution towards minimizing \Cref{eq:mincost}
    \State $\mathcal{P}_0 \gets \{\{v_1\},\dots,\{v_n\}\}$ \Comment{Initial node clusters}
    \State $\mathcal{Q}_0 \gets \{\{h_1\},\dots,\{h_m\}\}$ \Comment{Initial hyperedge clusters}
    \State $c_0 \gets \Call{computeCost}{\mathcal{P}_0, \mathcal{Q}_0, A}$ \Comment{Initial cost}
    \State $t \gets 0$, $D \gets A$ \Comment{Density matrix}
    \Repeat
        \State $\mathrm{improv} \gets \textbf{False}$, $c_{t+1} \gets c_t$
        \State $\mathcal{Q}_{t+1} \gets
	\Call{mergeClusters}{\mathcal{Q}_t, A, D, r, b, |\mathcal{P}_t|, \textbf{False}, t}$
        \State $\tilde{c} \gets \Call{computeCost}{\mathcal{P}_t, \mathcal{Q}_{t+1}, A, D}$
        \If{$c_t - \tilde{c} < \tau$}
            $\mathcal{Q}_{t+1} \gets \mathcal{Q}_t$
        \Else \label{line:merge1}
            $\: c_{t+1} \gets \tilde{c}$, $\mathrm{improv} \gets \textbf{True}$, 
        \EndIf
        \State $\mathcal{P}_{t+1} \gets
	\Call{mergeClusters}{\mathcal{P}_t, A, D, r, b, |\mathcal{Q}_t|, \textbf{True}, t}$
        \State $\tilde{c} \gets \Call{computeCost}{\mathcal{P}_{t+1}, \mathcal{Q}_{t+1}, A, D}$
        \If{$c_t - \tilde{c} < \tau$}
            $\mathcal{P}_{t+1} \gets \mathcal{P}_t$
        \Else \label{line:merge2}
            $\: c_{t+1} \gets \tilde{c}$, $\mathrm{improv} \gets \textbf{True}$, 
        \EndIf
        \State $(\mathcal{P}^\ast, \mathcal{Q}^\ast) \gets (\mathcal{P}_{t+1}, \mathcal{Q}_{t+1})$, $t \gets t + 1$
        \State Update density matrix $D$
    \Until{\textbf{not} $\mathrm{improv}$}\label{line:stopping} \Comment{No merging made a cost improvement}
    \State \Return $\Call{createSummary}{\mathcal{P}^\ast, \mathcal{Q}^\ast, D}$\label{line:return}
    \end{algorithmic}
\end{algorithm}

At each iteration, the algorithm alternates between merging hyperedge clusters and node clusters.
Candidate merges are identified via locality-sensitive hashing (LSH)
(see paragraph \textbf{Incremental cost updates})
and their potential cost improvement is
evaluated using local incremental updates based on our relaxed
objective. 
If the merges
reduce the total cost by at least a threshold $\tau$,
then they are accepted;
otherwise, they are discarded.
The process continues until no merge identified reduces the cost by more
than $\tau$ (\Cref{alg:algorithm} line~\ref{line:stopping}).
Parameter $\tau$ controls the tradeoff between runtime and quality:
smaller $\tau$ values yield more compact summaries
(more iterations), whereas larger values produce faster but larger summaries.

\begin{algorithm}[!ht]
    \caption{\textsc{computeCost}}
    \label{alg:computecost}
    \begin{algorithmic}[1]
    \Require Node Clusters $\mathcal{P}$, Hyperedge Clusters $\mathcal{Q}$
    \Require Hypergraph Incidence Matrix $A$, Density Matrix $D$
    \Ensure Cost of the summary obtained from $\mathcal{P}$ and $\mathcal{Q}$
    \State $\superedges \gets$ superhyperedges given $\mathcal{P}$ and $\mathcal{Q}$, computed via $D$
    \State $c_a \gets |\superedges|$ \Comment{Term \textcolor{red}{(a)} \Cref{eq:mincost}}
    \State $c_b \gets \sum_{r \in \superedges}|r|$ \Comment{Term \textcolor{red}{(b)} \Cref{eq:mincost}}
    \State $c_c^+ \gets 0$, $c_c^- \gets 0$
    \For{$r \in \superedges$}
        \State $Q \gets$ hyperedge cluster corresponding to $r$
        \State $c_c^- \gets c_c^- + \sum_{P \in r}\sum_{v \in P}\sum_{h \in Q}
	\left(1 - \mathds{1}_{A[v,h]}\right)$
        \State $c_c^+ \gets c_c^+ + \sum_{P \in \supernodes \setminus r}\sum_{v \in P}\sum_{h \in Q}\mathds{1}_{A[v,h]}$ 
    \EndFor
    \State $c_c \gets c_c^+ + c_c^-$ \Comment{Term \textcolor{red}{(c)} \Cref{eq:mincost}}
    \State \Return $c_a + c_b + c_c$
    \end{algorithmic}
\end{algorithm}

\spara{Cost computation.}
Given that in each step of our algorithm we have to evaluate the cost of
potential merges, we need to perform this cost computation efficiently.
\Cref{alg:computecost}
evaluates the cost of the summary implied by the current cluster
assignments, after the tentative merges suggested by
\Cref{alg:mergeclusters}.
Because recomputing the full correction table would be expensive, we
exploit the density matrix $D$ to infer supernode–superhyperedge
relations and compute correction counts efficiently.
We recall that corrections of type $c^-$ correspond to cases where a
node $v$ in a supernode $P$ included in a superhyperedge $r$ does
\emph{not} belong to one of the hyperedges represented by $r$.
Conversely, corrections of type $c^+$ count nodes $v$ in a supernode $P$
not included in $r$ but belonging to a hyperedge covered by $r$.
These quantities are computed efficiently via sparse matrix operations.

\begin{algorithm}[!th]
    \caption{\textsc{mergeClusters}}
    \label{alg:mergeclusters}
    \begin{algorithmic}[1]
    \Require Clusters $\mathcal{C}$, Hypergraph Incidence Matrix $A$
    \Require Density Matrix $D$, LSH parameters $r, b$
    \Require Number of clusters in the other dimension $k$ 
    \Require Cluster rows or columns? $\mathrm{clusterRows}$, Iteration $t$
    \Ensure New cluster assignment $\tilde{\mathcal{C}}$ for $\mathcal{C}$
    \If{$\mathrm{clusterRows}$} 
        $Z \gets D^\intercal$ \textbf{else} $Z \gets D$
    \EndIf
    \State $\ell \gets |\mathcal{C}|$, $S \gets$ matrix in $\mathbb{R}^{rb \times \ell}$\label{line:start_LSH}
    \If{$t = 1$} \Comment{Create min-hash signatures}
        \For{$i = 1$ to $rb$}
            \State $\pi_i \gets$ random permutation of $\{1,\dots,n\}$
            \For{$j = 1$ to $\ell$}
                $S[i][j] \gets \min_{v \in \Gamma_j} \pi_i(v)$
            \EndFor
        \EndFor
    \Else \Comment{Create random-projection signatures}
        \For{$i = 1$ to $rb$}
            \State $\mathrm{rnd}_i \gets$ random hyperplane in $\mathbb{R}^{k \times 1}$
            \For{$j = 1$ to $\ell$}
                $S[i][j] \gets \mathrm{sign}(\mathrm{rnd}_i \cdot Z[:][j])$
            \EndFor
        \EndFor
    \EndIf
    \For{$h = 1$ to $b$} \Comment{Generate hash tables}
        \For{$j = 1$ to $\ell_t$}
            $\Call{hash}{S[(h-1)r+1:hr][j]}$
        \EndFor
    \EndFor
    \State Group clusters in $\mathcal{C}$ hashed to $\geq 1$ same bucket in all tables\label{line:end_LSH}
    \For{each group $g$} \label{start:merging}
        \While{more merges happen}
            \State $C_a \gets$ random element from $g$
            \For{each other cluster $C \in g$}
                \State $c^* \gets \Call{computeDeltaCost}{C, C_a}$
                \If{$c^* \leq 0$}
                    $C_a \gets C_a \cup C$
                \Else
                    $\:\tilde{\mathcal{C}} \gets \tilde{\mathcal{C}} \cup \{C\}$
                \EndIf
            \EndFor
            \State $\tilde{\mathcal{C}} \gets \tilde{\mathcal{C}} \cup \{C_a\}$
        \EndWhile
    \EndFor \label{end:merging}
    \State \Return $\tilde{\mathcal{C}}$
    \end{algorithmic}
\end{algorithm}

\spara{Merging process.}
The merging procedure (shown in \Cref{alg:mergeclusters})
follows the same logic as in
\leman \cite{gao2014fast}, but is adapted to our storage-aware cost and
lossless summarization objective.
The same procedure is used for both node and hyperedge clusters,
alternating between the two until no further cost-reducing merge can
be found.

Each iteration may perform multiple merges.
To identify potential candidates, we employ LSH with parameters $r$
(signature size) and $b$ (number of hash tables), which efficiently
groups clusters with similar structural profiles into buckets. Each
bucket corresponds to a group of clusters that are likely to yield good
merges.
For each group $g$, we randomly select one cluster $C_a$ as the
\emph{anchor}. Each other cluster $C \in g$ is tentatively merged with the current anchor $C_a$.
This decision is made by calling the function \textsc{computeDeltaCost},
which computes the incremental change in cost by merging $C$ and~$C_a$.
If the result is negative, the merge would reduce the total cost, so we
merge $C$ and~$C_a$.
Clusters that do not yield improvement are added
directly to the output $\tilde{\mathcal{C}}$.

Once all candidates in group $g$ are processed, the anchor $C_a$ is
added to $\tilde{\mathcal{C}}$. After all groups are examined, the
procedure returns the candidate new clustering $\tilde{\mathcal{C}}$,
which is returned to \Cref{alg:algorithm} to determine whether to accept the merges.

\spara{Incremental cost updates.}
Our algorithm
must repeatedly evaluate
how the objective function changes after merging two clusters.
Recomputing the entire summary after each tentative merge would be
computationally infeasible.
Therefore, we implement an incremental update rule that computes the
variation in the objective efficiently using the
density matrix $D$ and the relaxed storage-aware cost function.

Let $C_1$ and $C_2$ be the two clusters to be merged.
We first identify the sets of clusters in the \emph{other dimension}
associated with them in the summary hypergraph implied by the current
clustering.
If $C_1$ and $C_2$ are node clusters, these sets correspond to the
hyperedge clusters whose superhyperedges would contain them.
If $C_1$ and $C_2$ are hyperedge clusters, the sets correspond to the
node clusters that their resulting superhyperedges would include.
These sets, denoted $\mathcal{B}_1 \doteq \big\{B_{i_1}, \dots, B_{i_{k_1}}\big\}$
and $\mathcal{B}_2 \doteq \big\{B_{j_1}, \dots, B_{j_{k_2}}\big\}$, and
the set $\mathcal{B}_{12}$ associated to
the cluster obtained by merging $C_1$ and $C_2$, are computed
directly from $D$ using sparse matrix operations.


We then define (for $i,j\in\{1,2\}, i\ne j$):
$\tilde{\mathcal{B}}_i \doteq \mathcal{B}_i \setminus \mathcal{B}_j$ to be the clusters only in $\mathcal{B}_i$,
and
$\tilde{\mathcal{B}}_{12} \doteq \mathcal{B}_1 \cap \mathcal{B}_2$ to be the clusters that are in both
$\mathcal{B}_1$ and $\mathcal{B}_2$.
The changes to the cost terms are as follows:

\emph{Term (a).}
It decreases by $1$ if $C_1$ and $C_2$ are hyperedge clusters (fewer
superhyperedges), unchanged otherwise (number of superhyperedges
remains the same).

\emph{Term (b).}
It decreases by $|\tilde{\mathcal{B}}_{12}|$, as shared clusters will
appear once in the resulting summary, and by $1$ for each cluster in
$(\tilde{\mathcal{B}}_1 \cup \tilde{\mathcal{B}}_2) \setminus
\mathcal{B}_{12}$, because that cluster will not be present after merging.

\emph{Term (c).}
For each cluster $B$ in $\mathcal{B}_{12} \cap \tilde{\mathcal{B}}_1$ (resp. $C_{12} \cap \tilde{C}_2$), it:
\begin{squishlist}
    \item increases by $1$ for each $0$ in the submatrix defined by $B
    \times C_2$ (resp. $B \times C_1$) as they correspond to $c^-$
    corrections that we must consider now that $\mathcal{B}_{12}$
    includes $B$; and
    \item decreases by $1$ for each $1$ in the submatrix defined by $B
    \times C_2$ (resp. $B \times C_1$) as they correspond to $c^+$
    corrections associated with $C_2$ (resp. $C_1$) because $B$ was not
    part of $\mathcal{B}_2$ (resp. $\mathcal{B}_1$).
\end{squishlist}

For each cluster $B$ in $\tilde{\mathcal{B}}_1 \setminus \mathcal{B}_{12}$
(resp. $\tilde{\mathcal{B}}_2 \setminus \mathcal{B}_{12}$), it:
\begin{squishlist}
  \item increases by $1$ for each $1$ in the submatrix defined by $B
  \times C_1$ (resp. $B \times C_2$) as they correspond to $c^+$
  corrections that we must consider given that $\mathcal{B}_{12}$ does not
  include $B$ while $\mathcal{B}_1$ (resp. $\mathcal{B}_2$) included it;
  and
  \item decreases by $1$ for each $0$ in the submatrix defined by $B
  \times C_1$ (resp. $B \times C_2$) as they correspond to $c^-$
  corrections associated with $C_1$ (resp. $C_1$) because $B$ was not part
  of $\mathcal{B}_1$ (resp. $\mathcal{B}_1$).
\end{squishlist}

To ensure scalability, these updates are computed using sparse matrix
operations on the relevant submatrices rather than on individual
entries.

\spara{Post-processing for summary generation.}
Unlike co-clustering, which outputs disjoint row and column clusters,
\our must produce a summary hypergraph and its
correction table.
Given the node and hyperedge clusters produced by the iterative
process, we construct the output summary in the following three steps. 

\noindent \textbf{(1)} Each row cluster is mapped to a supernode, and each column
  cluster to a superhyperedge. 
  
\noindent \textbf{(2)} For each superhyperedge $j$ and each supernode $i$, we include
  the supernode in the superhyperedge if $D[i,j] \geq 0.5$. This
  ensures that the cost associated with storing the association and
  the corrections for the zero entries in that submatrix is lower than
  the cost of storing the corrections for the ones in the
  submatrix.

\noindent \textbf{(3)} During the merge process, we maintain a mapping between the
  current cluster assignments and the evolving summary, which is
  required to compute incremental costs. However, because merges are
  performed in blocks of similar clusters, two superhyperedges may
  become identical at later stages; in this case, we merge them and
  update the associated corrections and weights accordingly.

\subsection{Complexity Analysis}
\label{sec:app-complexity-analysis}
The overall complexity of \our depends on three components: computing
the cost of the current summary (\Cref{alg:computecost}), identifying
merge candidates (\Cref{alg:mergeclusters}~lines
\ref{line:start_LSH}-\ref{line:end_LSH}), and performing merges
(\Cref{alg:mergeclusters}~lines \ref{start:merging}-\ref{end:merging}).

\spara{Cost evaluation.} 
\Cref{alg:computecost} examines all node–hyp\-er\-edge cluster pairs. If the
summary currently has $k$ supernodes and $\ell$ superhyperedges, each
containing on average $n/k$ nodes and $m/\ell$ hyperedges, the
complexity is $\mathcal{O}(nm)$, which is linear in the number of
nonzero entries of $A$ when implemented sparsely.

\spara{Merge candidate generation.}
\Cref{alg:mergeclusters} builds LSH signatures with cost
$\mathcal{O}((k+\ell)rb)$ and performs similarity grouping with
$\mathcal{O}(k\ell rb)$ hashing operations. Within each group, we
evaluate the incremental delta-cost; this dominates the step and yields
$\mathcal{O}(k^2\ell^2)$ in the worst case, though in practice, much
lower due to sparse updates.

\spara{Overall runtime.}
The main loop executes until no merge reduces the cost by more than the threshold $\tau$. 
If the algorithm converges in $T$ iterations, the total cost is
$\mathcal{O}(T(nm + (k+\ell)rb + k^2\ell^2))$.
In practice, $T$ is small (at most $23$ in our experiments), and $k,\ell \ll n,m$, yielding near-linear scaling in the number of hyperedges.
\section{Experimental Evaluation}\label{sec:exp}

Our experimental evaluation assesses the effectiveness of the proposed
lossless hypergraph-summarization approach. We address three questions:
\begin{squishlist}
    \item[\textbf{Q1}] How compact are our summaries compared to those generated by competing approaches?
    \item[\textbf{Q2}] How well do the summaries preserve query accuracy and downstream analysis results?
    \item[\textbf{Q3}] What is the computational cost of summary construction and algorithm execution on them?
\end{squishlist}

We evaluate these questions using diverse real-world hypergraphs, comparing against multiple baselines and metrics.

\spara{Datasets.}
We consider twelve real-world datasets taken from prior
work~\cite{akoglu2015graph,benson2018simplicial}.
Their characteristics are reported in
\Cref{tbl:datasets}.
\textsc{senate} and \textsc{house} are hypergraphs generated from bill co-sponsorship relations between legislators from the US Senate and the US House of Representatives;
\textsc{NDC-classes} are sets of classifications applied to drugs;
\textsc{dblp} models academic papers and their commonly used terms relations;
\textsc{classic} models documents and the words they contain;
\textsc{polblog} models political blogs and the words they use;
\textsc{enron} are sets of email addresses in emails;
\textsc{primary-school} and \textsc{high-school} are contact hyper-networks between students at a primary and a high school;
\textsc{NDC-substances} are sets of substances making up drugs;
\textsc{tags-math} are sets of tags applied to questions on \url{math.stackexchange.com};
and \textsc{DAWN} are sets of drugs used by patients recorded in emergency room visits.

\begin{table}[!ht]
    \centering
    \caption{Dataset characteristics: number of nodes, number of hyperedges, mean hyperedge size, max hyperedge size, mean node hyper-degree, and max node hyper-degree.}
    \label{tbl:datasets}
    \vspace{-3mm}
    \resizebox{\columnwidth}{!}{
        \begin{tabular}{lrrrrrr}
        \toprule
        \textbf{Dataset} & $|\nodes|$ & $|\edges|$ & $\hat{d}$ & $d_{\max}$ & $\hat{\hdegrf}$ & $\hdegrf_{\max}$ \\
        \midrule
        \textsc{senate} & 106 & 695 & 57.197 & 97 & 375.019 & 484 \\
        \textsc{NDC-classes} &	1161 &	1090 & 5.972	& 39 & 5.606 &	222 \\
        \textsc{dblp} & 1227 & 1227 & 15.702 & 101 & 15.702 & 101 \\
        \textsc{house} & 451 & 1646 & 304.740 & 433 & 1112.199 & 1357 \\
        \textsc{classic} & 3891 & 4303 & 40.982 & 578 & 45.322 & 172 \\
        \textsc{polblog} & 362 & 5895 & 131.784 & 362 & 2146.050 & 5068 \\
        \textsc{enron} & 143 & 10885 & 2.472 & 37 & 188.210 & 1327 \\
        \textsc{primary-school} & 242 & 106879 & 2.096 & 5 & 925.611 & 2234 \\
        \textsc{NDC-substances} & 5556 & 112919 & 2.015 & 187 & 40.946 & 6693 \\
        \textsc{high-school} & 327 & 172035 & 2.050 & 5 & 1078.648 & 4495 \\
        \textsc{tags-math} & 1629 & 822059 & 2.192 & 5 & 1105.977 & 71046 \\
        \textsc{DAWN} & 2558 & 2272433 & 1.583 & 16 & 1406.096 & 479242 \\
        \bottomrule
        \end{tabular}
    }
\end{table}

\spara{Evaluation metrics.}
We evaluate all methods along three verticals: running
time, space efficiency, and accuracy in query answering.
Running time includes both summary construction and downstream analyses.
Space efficiency is measured by (1) disk space relative to the original
hypergraph and (2) improvement under our storage-cost measure; the
former reflects practical space usage under our CSV encoding, whereas
the latter is encoding-independent and highlights the intrinsic
information
captured by the summary.
Accuracy is measured by agreement between results on
the original hypergraph and its summary, using normalized discounted cumulative gain (NDCG), Jaccard index, MAE, and MSE.

\spara{Baselines.}
We compare our approach against several co-clustering algorithms.
\leman~\cite{gao2014fast} is the framework underlying our
algorithm \our.
SC~\cite{dhillon2001co} is a spectral co-clustering method based on
bipartite graph partitioning, which simultaneously groups rows and
col\-umns by exploiting the normalized-cut criterion.
SB~\cite{kluger2003spectral} is a spectral biclustering algorithm that
performs eigenvalue decomposition of normalized incidence matrices to
identify coherent blocks.
\cc~\cite{ienco2013parameter} is a parameter-free co-clustering
algorithm for star-structured heterogeneous data, which
automatically infers the number of clusters.
\tcc~\cite{battaglia2024fast} is a prototype-based, parameter-free co-clustering
algorithm that achieves state-of-the-art
scalability and quality on large datasets.

We also include two random baselines: rand-eqS and rand-randS.
Given parameters $k$ and $\ell$, the first generates $k$ node partitions
and $\ell$ hyperedge partitions of similar size, and the second assigns
random partition sizes before assigning nodes and hyperedges to the partitions.
In our experiments, $k$ and $\ell$ are set to the mean number of node and
hyperedge clusters found by the parameter-free methods.
Superedges induced by identical supernode sets are
collapsed, so the effective number of superedges may vary.

We exclude methods that summarize graph projections of
hypergraphs, as it is unclear how to recover a hypergraph summary
from the projection summary, and how to group the original hyperedges into superhyperedges.
We also exclude hypergraph clustering
methods~\cite{li2023efficient,purkait2016clustering,bulo2009game}, as
they partition nodes but not hyperedges, whereas both are required for our task.

All experiments were run on a $16$-core Ubuntu $24.04$ system with
$125$ GB RAM using Python 3.13. All algorithms are executed $5$ times,
and we report average results.
For \cc and \tcc, we used the default hyperparameter settings.
For \leman and \our, we set $r=20$ and $b=5$.
For \leman, we set $\tau = \text{1e-6}$, and for \our we
set $\tau = 1$.
More information on the hyperparameter tuning for \our is in \Cref{sec:app-hyperparameters}.
Briefly, the sensitivity analysis shows that the hashing parameters
$(r,b)$ mainly affect runtime--larger values slow the method slightly but
yield only minor and inconsistent cost improvements.
The stopping threshold $\tau$ governs the runtime–accuracy tra\-deoff:
smaller values allow more merges but increase runtime, whereas larger values terminate earlier.
Overall, performance is stable across reasonable parameter choices.

\noindent
The source code is available at \url{https://github.com/lady-bluecopper/HyDRA}.

\subsection{Summarization Quality}\label{sec:summ_quality}
We evaluate the algorithms in terms of the quality of their summaries, considering space savings with respect to the original
hypergraph, cost improvement, and runtime. The results of this analysis
are reported in \Cref{tbl:summarization}.

\emph{Random baselines.}
As expected, the random-part\-it\-ioning baselines are the fastest methods,
as they do not optimize any objective function. However, because
nodes and hyperedges are grouped arbitrarily, they consistently
achieve the worst performance in terms of both space savings and cost
improvement.

\emph{Spectral clustering baselines.}
SC and SB rely on spectral methods to co-cluster the incidence matrix.
SC minimizes the normalized cut using SVD followed by $k$-means, which
makes it relatively efficient. SB, in contrast, requires
bistochastization (iterative row/column normalization until
convergence), which can lead to substantially higher runtimes
(especially for larger values of required node clusters $k$), though
sometimes with slightly better quality. A key limitation of both methods
is their reliance on the number of node clusters as input, which
restricts their applicability.

\emph{Prototype-based co-clustering.}
The CC family relies on prototype-based optimization
strategies, which are fast and group rows and columns to maximize
statistical association (Goodman–Kruskal $\tau$). Although effective in
finding coherent biclusters, these methods are not designed to produce
lossless hypergraph summaries. As a result, their cost improvements are
systematically lower than those of the \leman family.

\emph{LSH-based variants.}
Both \leman and our proposed \our directly tackle a hard optimization
problem, solved greedily via iterative merges until convergence. This
explains their larger runtimes compared to the other families. However,
this additional cost pays off: these methods optimize storage-related
objectives--clustering cost and summary storage cost in \our--and
therefore achieve substantially better cost improvements and space
savings. Our algorithm, \our, consistently delivers the best results,
often by a large margin. The only exceptions are (1) the
\textsc{classic} dataset, which lacks a clear \emph{checkerboard} structure (i.e., when rows and columns of the incidence matrix can be reordered to reveal alternating dense and sparse blocks)
and thus resists compression by any algorithm, and (2) the
\textsc{polblog} dataset, where the two variants perform equally well.

\begin{table*}[!t]
\caption{Summarization results: average number of node and hyperedge
clusters, average cost improvement of the summary, time required to
find the summary, and space saved by storing the summary over the
input hypergraph. The symbol $-$ means that the algorithm did not
terminate within $3$ days of computation, and OOM means that it
failed because of memory errors.}
\label{tbl:summarization}
\resizebox{\linewidth}{!}{
\begin{tabular}{l|l|rrrrr||rrrrr|r}
\multicolumn{1}{l}{} & \multicolumn{1}{l}{\multirow{2}{*}{\textbf{Algorithm}}} & \multirow{2}{*}{\textbf{N Cl.}} & \multirow{2}{*}{\textbf{H Cl.}} & \multirow{2}{*}{\textbf{Time (s)}} & \textbf{Cost} & \multicolumn{1}{l}{\textbf{Space}} & \multirow{2}{*}{\textbf{N Cl.}} & \multirow{2}{*}{\textbf{H Cl.}} & \multirow{2}{*}{\textbf{Time (s)}} & \textbf{Cost} & \multicolumn{1}{l}{\textbf{Space}} & \\
 \multicolumn{1}{l}{} &  \multicolumn{1}{l}{} & & & & \textbf{Impr.} & \multicolumn{1}{l}{\textbf{Saved (\%)}} & & & & \textbf{Impr.} & \multicolumn{1}{l}{\textbf{Saved (\%)}} & \\
\cline{2-12}
\multirow{8}{*}{\rotatebox[origin=c]{90}{\textsc{senate}}}
& rand-eqS & 6.0 & 2.8 & 0.037 & 0.182 & 16.379 & 70.0 & 1.0 & 19.164 & 0.323 & 10.324 & \multirow{8}{*}{\rotatebox[origin=c]{90}{\textsc{primary-school}}} \\
& rand-randS & 6.0 & 3.2 & 0.034 & 0.182 & 16.724 & 70.0 & 122.2 & 8.391 & 0.324 & 10.425 & \\
& SC & 5.0 & 4.0 & 1.282 & 0.679 & 65.983 & 49.0 & 9.2 & 75.757 & 0.334 & 11.992 & \\
& SB & 6.0 & 4.8 & 20.961 & 0.728 & 71.453 & 70.0 & 20.2 & 665.206 & 0.336 & 12.154 & \\
& \cc & 2.0 & 2.0 & 0.427 & 0.611 & 60.684 & 12.6 & 3.6 & 59.599 & 0.331 & 11.507 & \\
& \tcc & 2.0 & 2.0 & 0.050 & 0.611 & 60.684 & 2.2 & 1.2 & 124.985 & 0.323 & 10.556 & \\
& \leman & 11.2 & 9.6 & 6.812 & 0.811 & 80.000 & 23.6 & 3.4 & 809.141 & 0.325 & 10.880 & \\
& \textbf{\our} & 15.0 & 15.6 & 8.078 & 0.816 & \textbf{80.513} & 242.0 & 3901.2 & 930.414 & 0.873 & \textbf{83.964} & \\
\cline{2-12}
\multirow{8}{*}{\rotatebox[origin=c]{90}{\textsc{dblp}}}
& rand-eqS & 25.0 & 1.0 & 0.142 & 0.060 & 2.381 & 45.0 & 1.0 & 0.611 & 0.288 & 7.619 & \multirow{8}{*}{\rotatebox[origin=c]{90}{\textsc{enron}}}\\
& rand-randS & 25.0 & 5.0 & 0.097 & 0.060 & 2.381 & 45.0 & 33.4 & 0.505 & 0.288 & 7.619 & \\
& SC & 25.0 & 2.6 & 11.216 & 0.060 & 1.190 & 40.0 & 10.4 & 18.300 & 0.344 & 14.667 & \\
& SB & 25.0 & 5.2 & 49.406 & 0.259 & 19.762 & 45.0 & 16.0 & 214.280 & 0.358 & 17.143 & \\
& \cc & 6.2 & 6.2 & 6.176 & 0.060 & 1.190 & 10.2 & 3.2 & 2.414 & 0.296 & 11.238 & \\
& \tcc & 5.0 & 5.0 & 6.034 & 0.060 & 1.190 & 4.0 & 1.2 & 6.699 & 0.289 & 10.667 & \\
& \leman & 4.8 & 1.4 & 107.801 & 0.078 & 2.857 & 22.8 & 6.0 & 56.335 & 0.318 & 14.476 & \\
& \textbf{\our} & 123.2 & 105.4 & 192.933 & 0.614 & \textbf{55.238} & 143.0 & 512.2 & 66.041 & 0.851 & \textbf{81.143} & \\
\cline{2-12}
\multirow{8}{*}{\rotatebox[origin=c]{90}{\textsc{house}}}
& rand-eqS & 10.0 & 1.0 & 1.352 & 0.522 & 51.335 & 681.0 & 1.0 & 40.282 & 0.332 & 8.333 & \multirow{8}{*}{\rotatebox[origin=c]{90}{\textsc{NDC-substances}}} \\
& rand-randS & 10.0 & 2.6 & 1.177 & 0.522 & 51.432 & 681.0 & 58.6 & 16.936 & 0.332 & 8.333 & \\
& SC & 8.0 & 6.2 & 3.348 & 0.818 & 81.185 & 681.0 & 331.4 & 396.168 & 0.331 & 8.333 & \\
& SB & 10.0 & 6.0 & 17.787 & 0.813 & 80.903 & 681.0 & 301.6 & 9975.606 & 0.483 & 30.192 & \\
& \cc & 2.0 & 2.0 & 0.391 & 0.681 & 68.370 & 5.2 & 1.0 & 279.392 & 0.332 & 8.333 & \\
& \tcc & 2.0 & 2.0 & 0.704 & 0.681 & 68.370 & 2.2 & 1.0 & 249.459 & 0.332 & 8.333 & \\
& \leman & 25.8 & 19.8 & 51.843 & 0.868 & \textbf{86.534} & 20.2 & 3.8 & 1557.116 & 0.336 & 8.333 & \\
& \textbf{\our} & 25.8 & 21.4 & 56.577 & 0.865 & 86.274 & 2697.8 & 2057.6 & 3035.240 & 0.727 & \textbf{61.898} & \\
\cline{2-12}
\multirow{8}{*}{\rotatebox[origin=c]{90}{\textsc{classic}}}
& rand-eqS & 649.0 & 1.0 & 9.055 & 0.024 & \textbf{0.475} & 52.0 & 1.0 & 0.094 & 0.143 & 0.000 & \multirow{8}{*}{\rotatebox[origin=c]{90}{\textsc{NDC-classes}}}\\
& rand-randS & 649.0 & 648.2 & 8.248 & 0.020 & \textbf{0.475} & 52.0 & 7.4 & 0.073 & 0.143 & 0.000 & \\
& SC & 483.6 & 218.2 & 45.238 & 0.022 & 0.356 & 52.0 & 18.0 & 3.902 & 0.141 & 0.000 & \\
& SB & 649.0 & 388.4 & 171.812 & 0.011 & -0.095 & 52.0 & 10.4 & 11.434 & 0.257 & 10.323 & \\
& \cc & 2.4 & 2.4 & 18.015 & 0.024 & 0.474 & 20.8 & 1.0 & 7.128 & 0.143 & 6.452 & \\
& \tcc & 2.8 & 2.8 & 11.560 & 0.024 & 0.474 & 2.6 & 1.0 & 2.538 & 0.143 & -2.837 & \\
& \leman & 1.4 & 1.0 & 974.750 & 0.024 & 0.474 & 3.6 & 1.0 & 61.027 & 0.143 & 6.452 & \\
& \textbf{\our} & 3881.2 & 4298.2 & 1865.653 & 0.003 & 0.356 & 182.2 & 194.6 & 140.601 & 0.539 & \textbf{39.355} & \\
\cline{2-12}
\multirow{8}{*}{\rotatebox[origin=c]{90}{\textsc{polblog}}}
& rand-eqS & 61.0 & 236.6 & 3.349 & 0.025 & 0.000 & 431.0 & 1.0 & 207.020 & 0.313 & 10.526 & \multirow{8}{*}{\rotatebox[origin=c]{90}{\textsc{tags-math}}}\\
& rand-randS & 61.0 & 350.2 & 3.060 & 0.050 & 3.704 & 431.0 & 348.8 & 66.679 & 0.314 & 10.526 & \\
& SC & 51.0 & 56.4 & 35.195 & 0.336 & 32.143 & 380.0 & 62.0 & 762.870 & 0.328 & 11.842 & \\
& SB & 61.0 & 53.2 & 53.993 & 0.349 & 34.286 & - & - & - & - & - & \\
& \cc & 2.0 & 2.0 & 1.050 & 0.187 & 17.857 & 8.4 & 1.2 & 164.259 & 0.314 & 10.526 & \\
& \tcc & 2.0 & 2.0 & 2.363 & 0.186 & 17.857 & 4.0 & 1.0 & 475.003 & 0.313 & 10.526 & \\
& \leman & 20.6 & 3.6 & 347.587 & 0.351 & \textbf{35.714} & 82.4 & 3.4 & 29925.654 & 0.316 & 10.789 & \\
& \textbf{\our} & 338.2 & 2113.6 & 1088.031 & 0.371 & \textbf{35.714} & 1629.0 & 13660.0 & 26967.858 & 0.723 & \textbf{64.211} & \\
\cline{2-12}
\multirow{8}{*}{\rotatebox[origin=c]{90}{\textsc{high-school}}}
& rand-eqS & 106.0 & 1.0 & 38.514 & 0.328 & 6.667 & 1314.0 & 1.0 & 2185.420 & 0.387 & -21.429 & \multirow{8}{*}{\rotatebox[origin=c]{90}{\textsc{DAWN}}}\\
& rand-randS & 106.0 & 128.2 & 21.556 & 0.328 & 6.667 & 1314.0 & 626.8 & 573.654 & 0.388 & -21.429 & \\
& SC & 98.2 & 43.4 & 82.402 & 0.433 & 25.000 & - & - & - & - & - & \\
& SB & 106.0 & 26.2 & 923.914 & 0.377 & 18.750 & - & - & - & - & - & \\
& \cc & 11.0 & 2.8 & 54.019 & 0.334 & 12.500 & OOM & OOM & OOM & OOM & OOM & \\
& \tcc & 2.0 & 1.0 & 256.071 & 0.328 & 12.500 & OOM & OOM & OOM & OOM & OOM & \\
& \leman & 85.6 & 30.0 & 1727.645 & 0.413 & 25.000 & 69.6 & 11.4 & 118648.163 & 0.439 & 25.333 & \\
& \textbf{\our}& 327.0 & 3062.8 & 981.360 & 0.950 & \textcolor{red}{\textbf{93.811}} & 2558.0 & 10947.6 & 55278.145 & 0.885 & \textbf{84.133} & \\
\cline{2-12}
\end{tabular}
}
\end{table*}

\subsection{Node and Hyperedge Cluster Size Distributions}

\begin{figure*}[ht]
    \centering
    \begin{subfigure}{.32\linewidth}
    \centering
        \includegraphics[width=\linewidth]{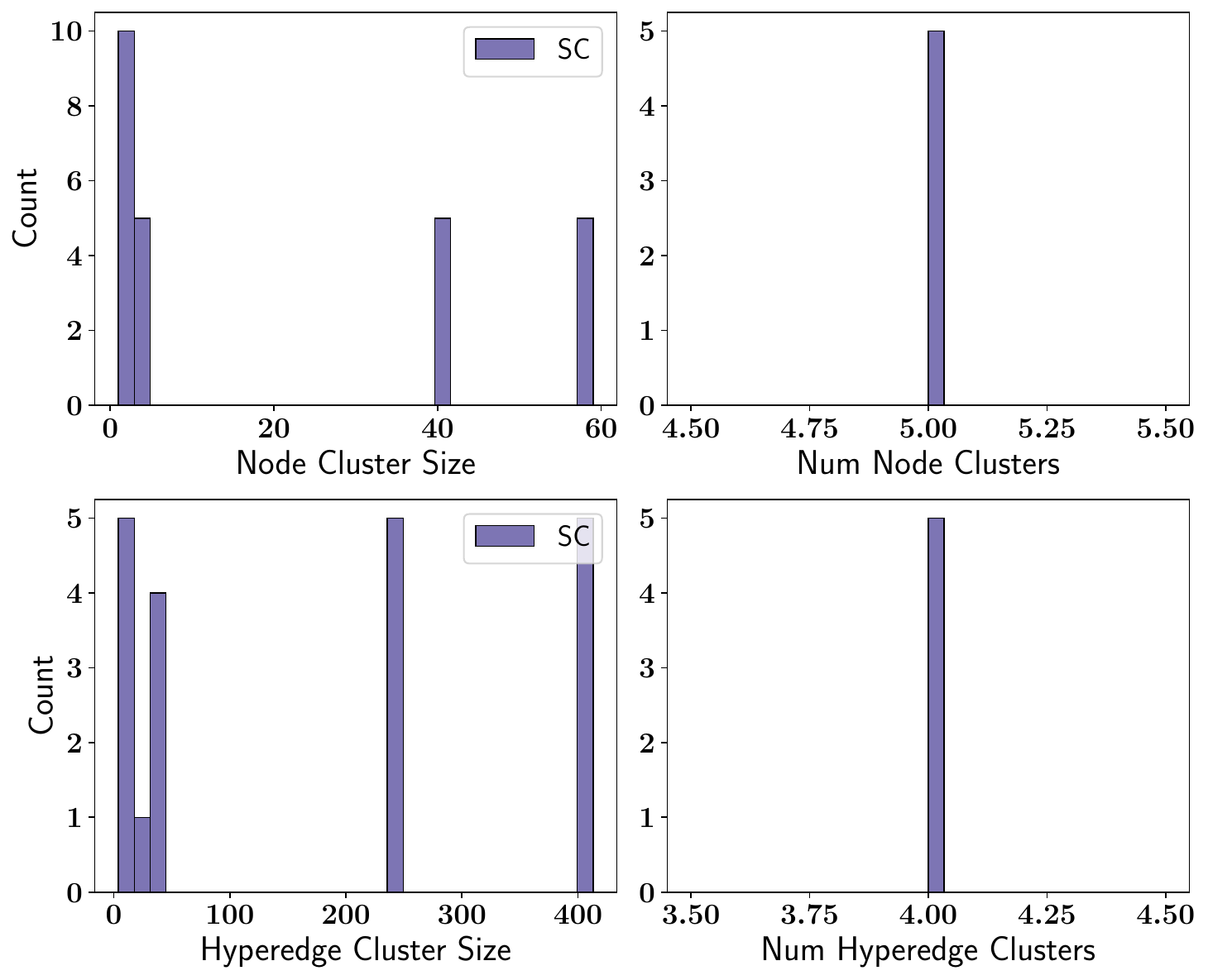}
        \caption{SC}
    \end{subfigure}
    \begin{subfigure}{.32\linewidth}
    \centering
        \includegraphics[width=\linewidth]{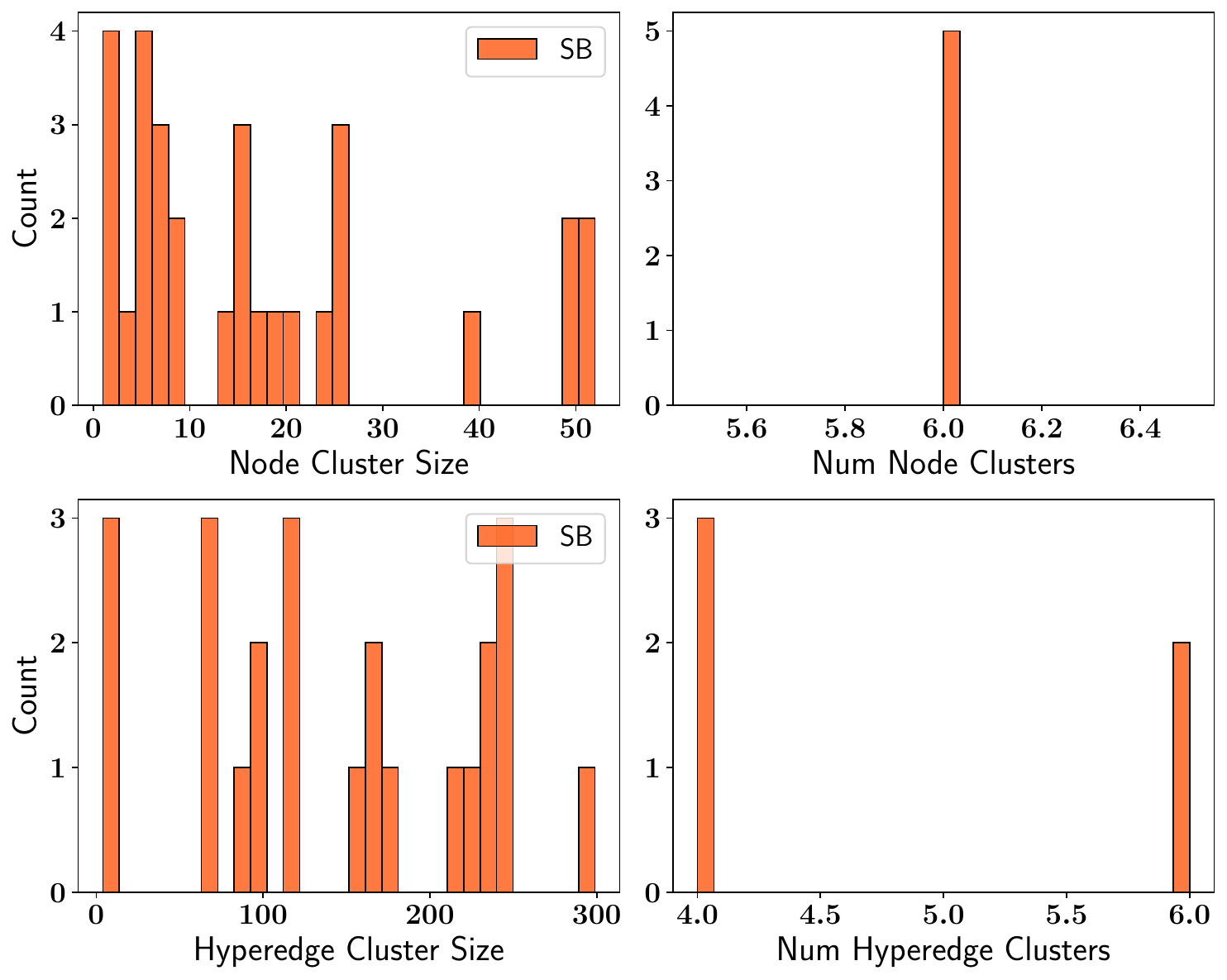}
        \caption{SB}
    \end{subfigure}
    \begin{subfigure}{.32\linewidth}
    \centering
        \includegraphics[width=\linewidth]{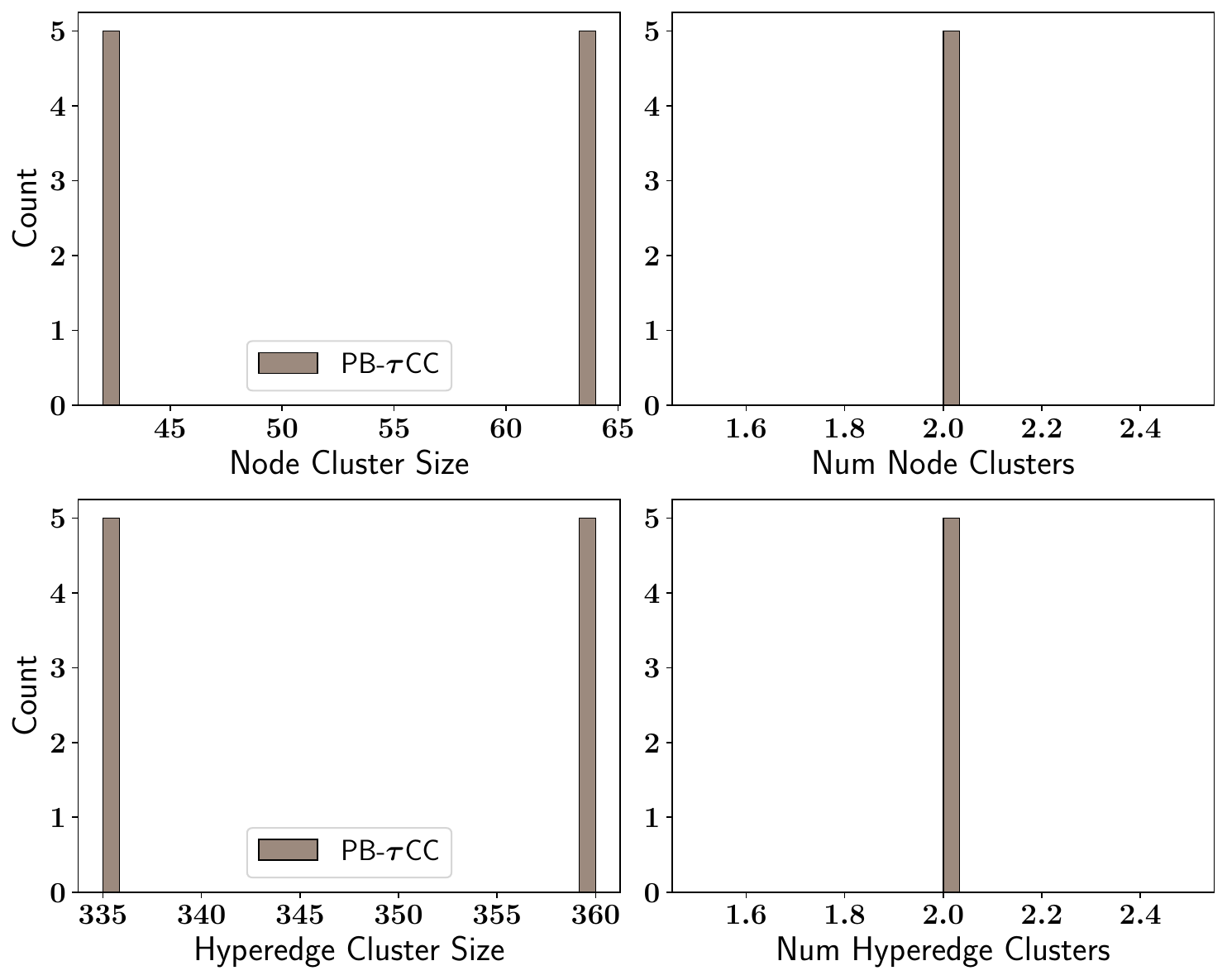}
        \caption{\cc}
    \end{subfigure}
    \begin{subfigure}{.32\linewidth}
    \centering
        \includegraphics[width=\linewidth]{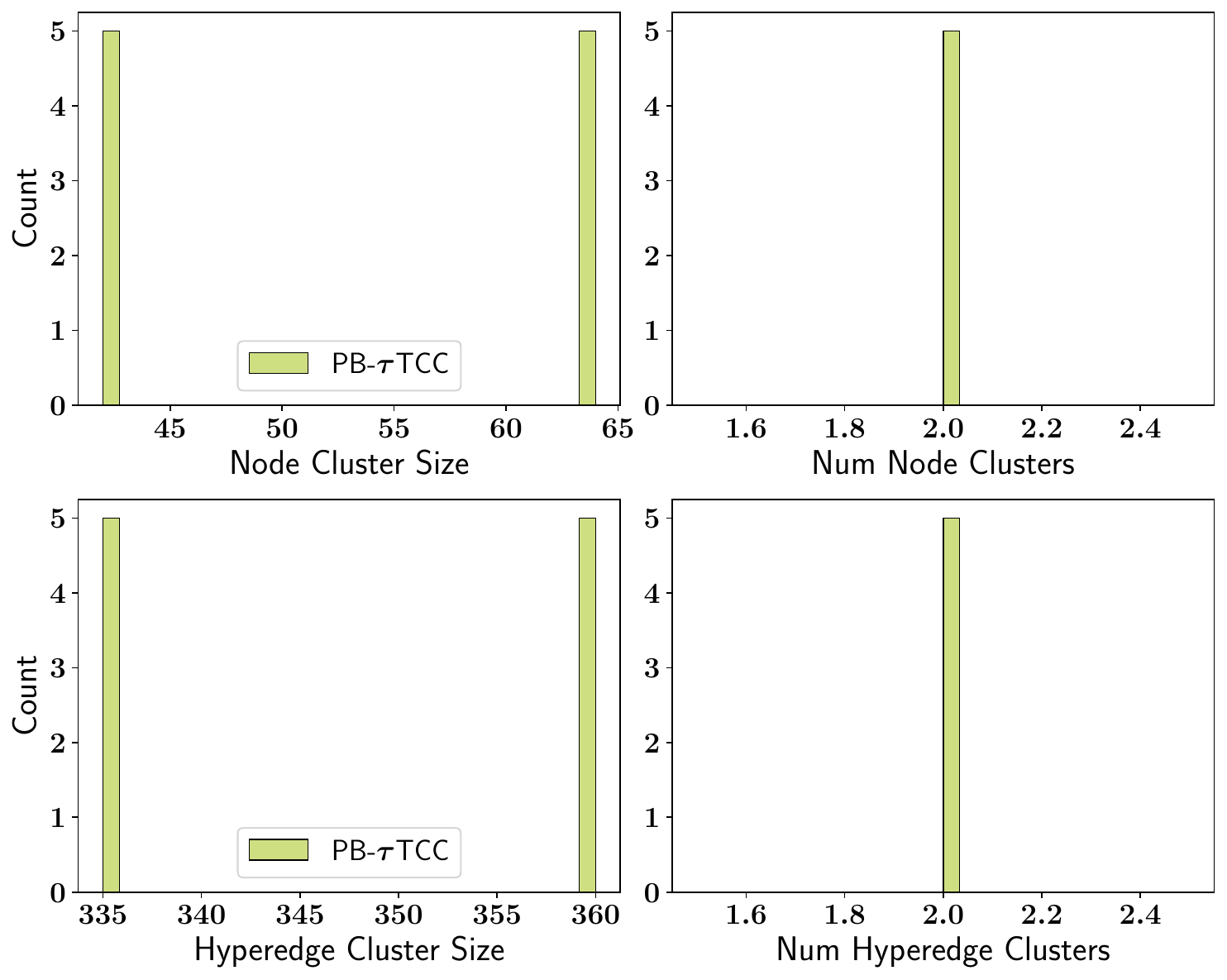}
        \caption{\tcc}
    \end{subfigure}
    \begin{subfigure}{.32\linewidth}
    \centering
        \includegraphics[width=\linewidth]{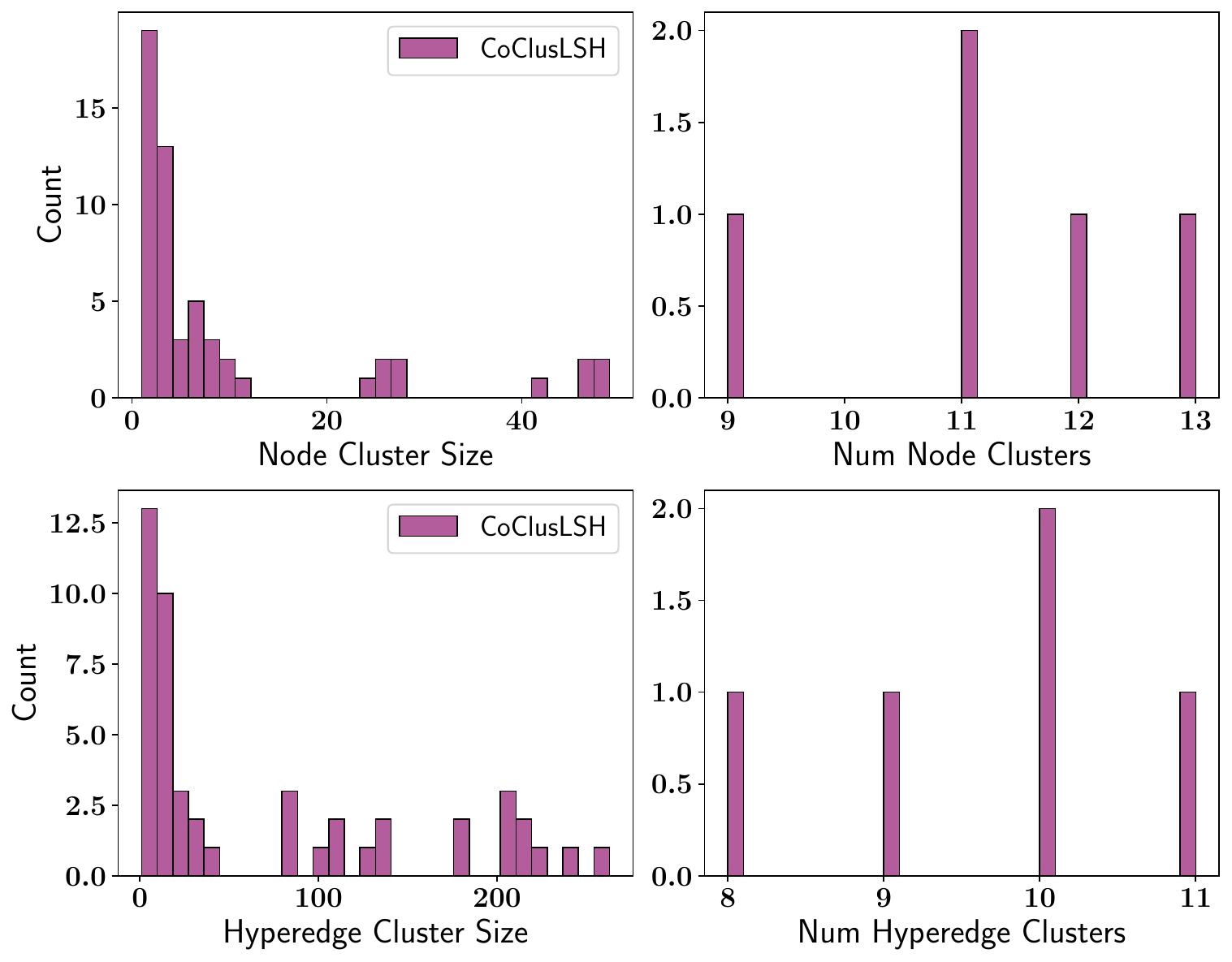}
        \caption{\leman}
    \end{subfigure}
    \begin{subfigure}{.32\linewidth}
    \centering
        \includegraphics[width=\linewidth]{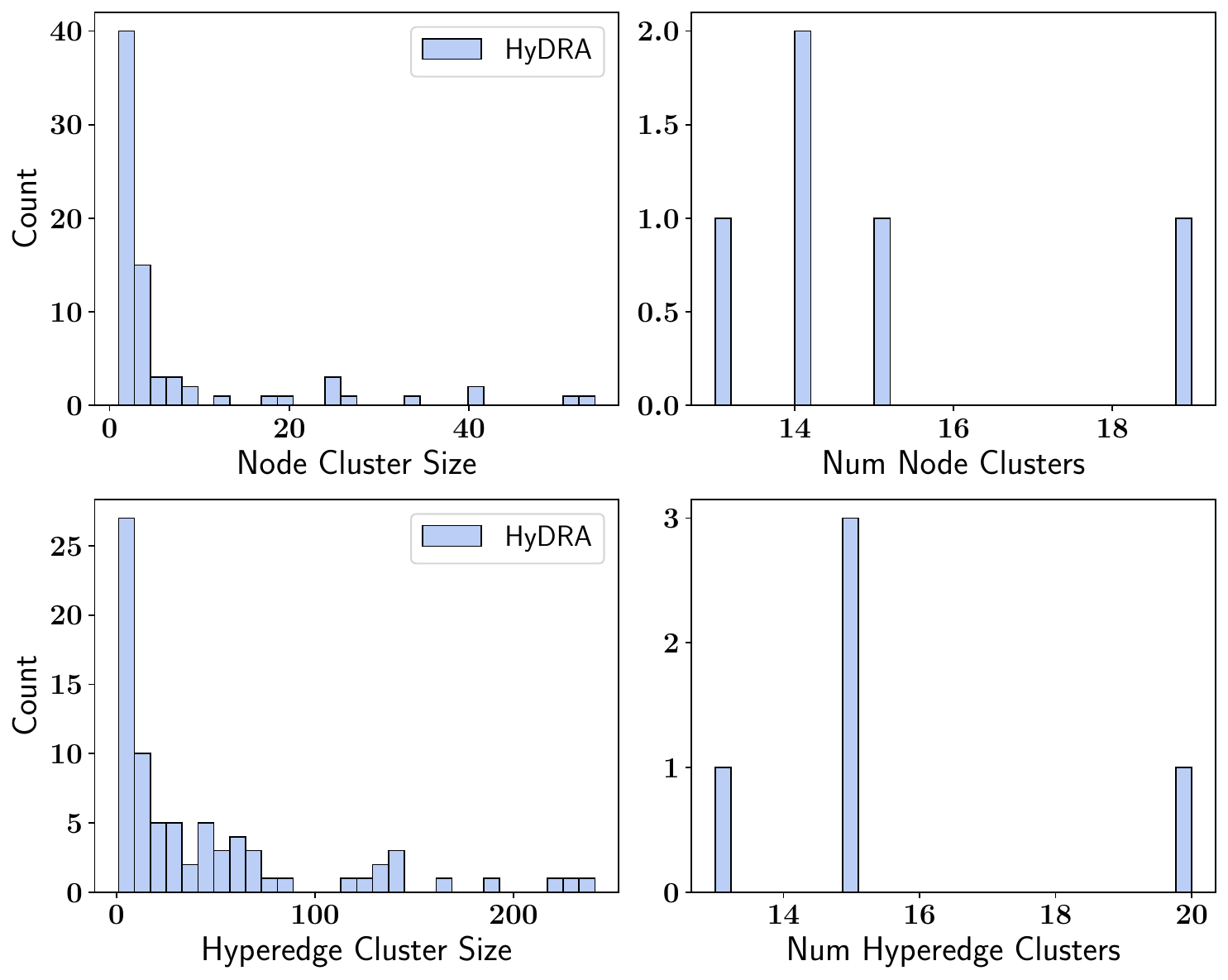}
        \caption{\our}
    \end{subfigure}
    \caption{Node and hyperedge cluster size distributions for summaries of \textsc{senate}.}
    \label{fig:clusters_senate}
\end{figure*}

\begin{figure*}[ht]
    \centering
    \begin{subfigure}{.32\linewidth}
    \centering
        \includegraphics[width=\linewidth]{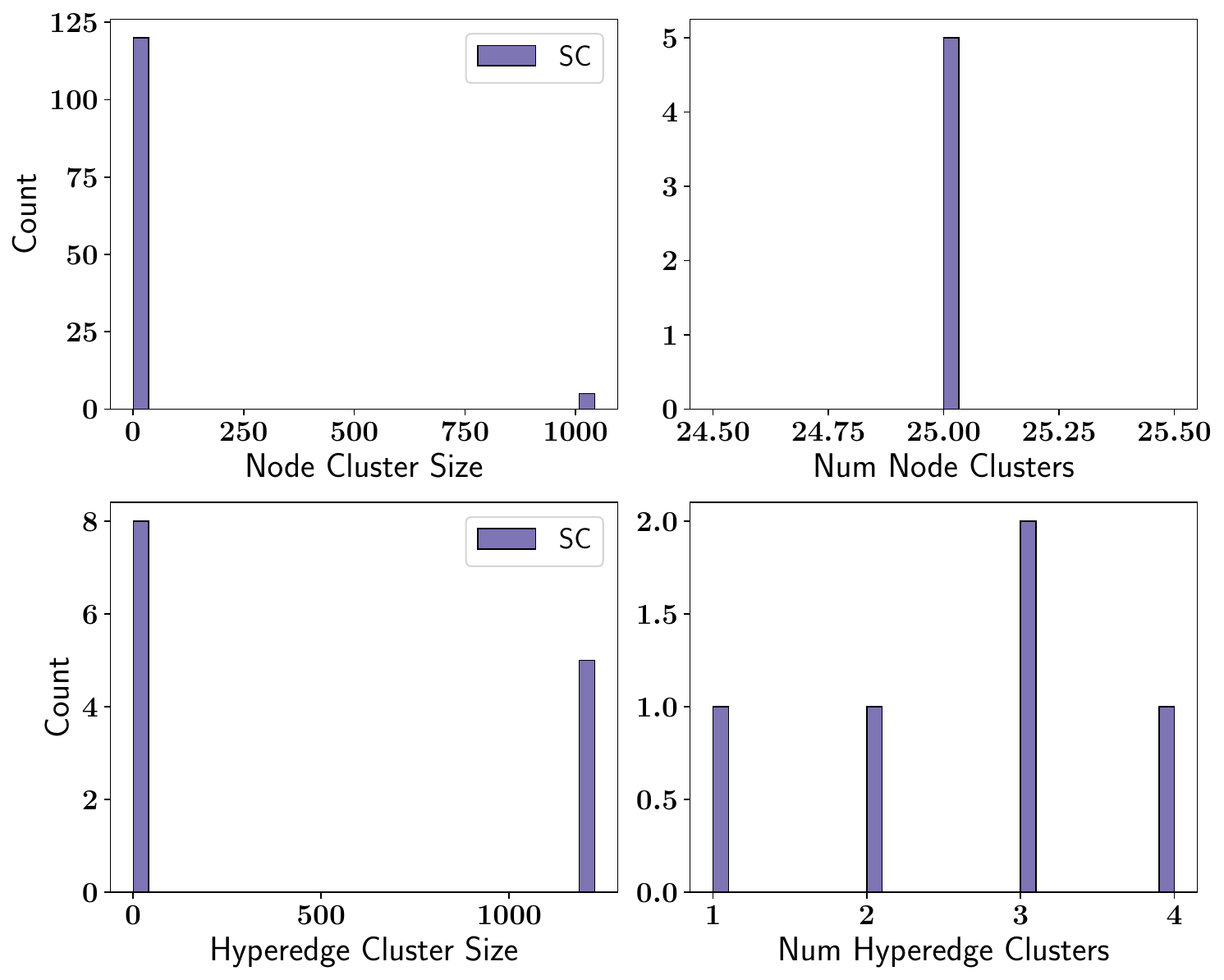}
        \caption{SC}
    \end{subfigure}
    \begin{subfigure}{.32\linewidth}
    \centering
        \includegraphics[width=\linewidth]{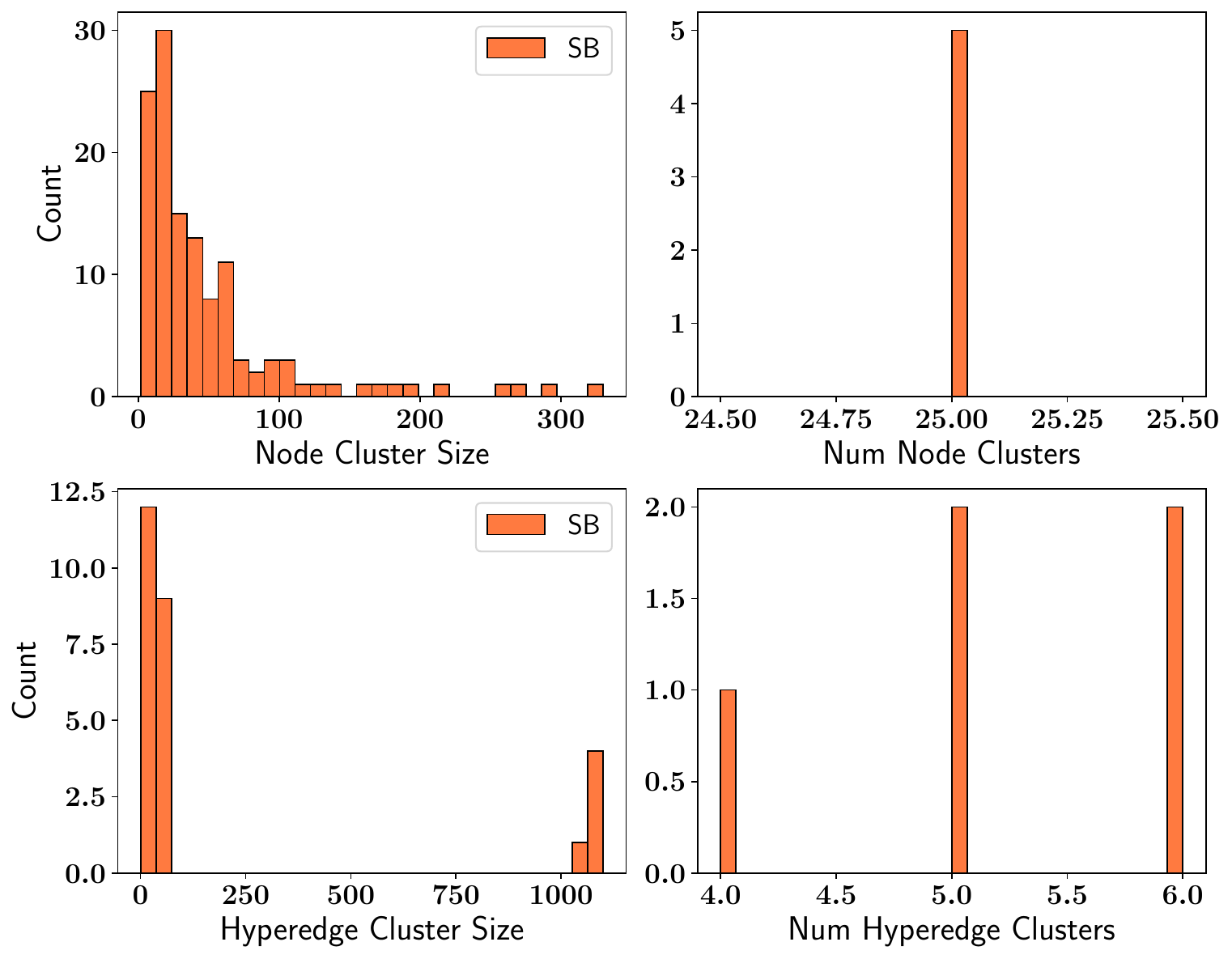}
        \caption{SB}
    \end{subfigure}
    \begin{subfigure}{.32\linewidth}
    \centering
        \includegraphics[width=\linewidth]{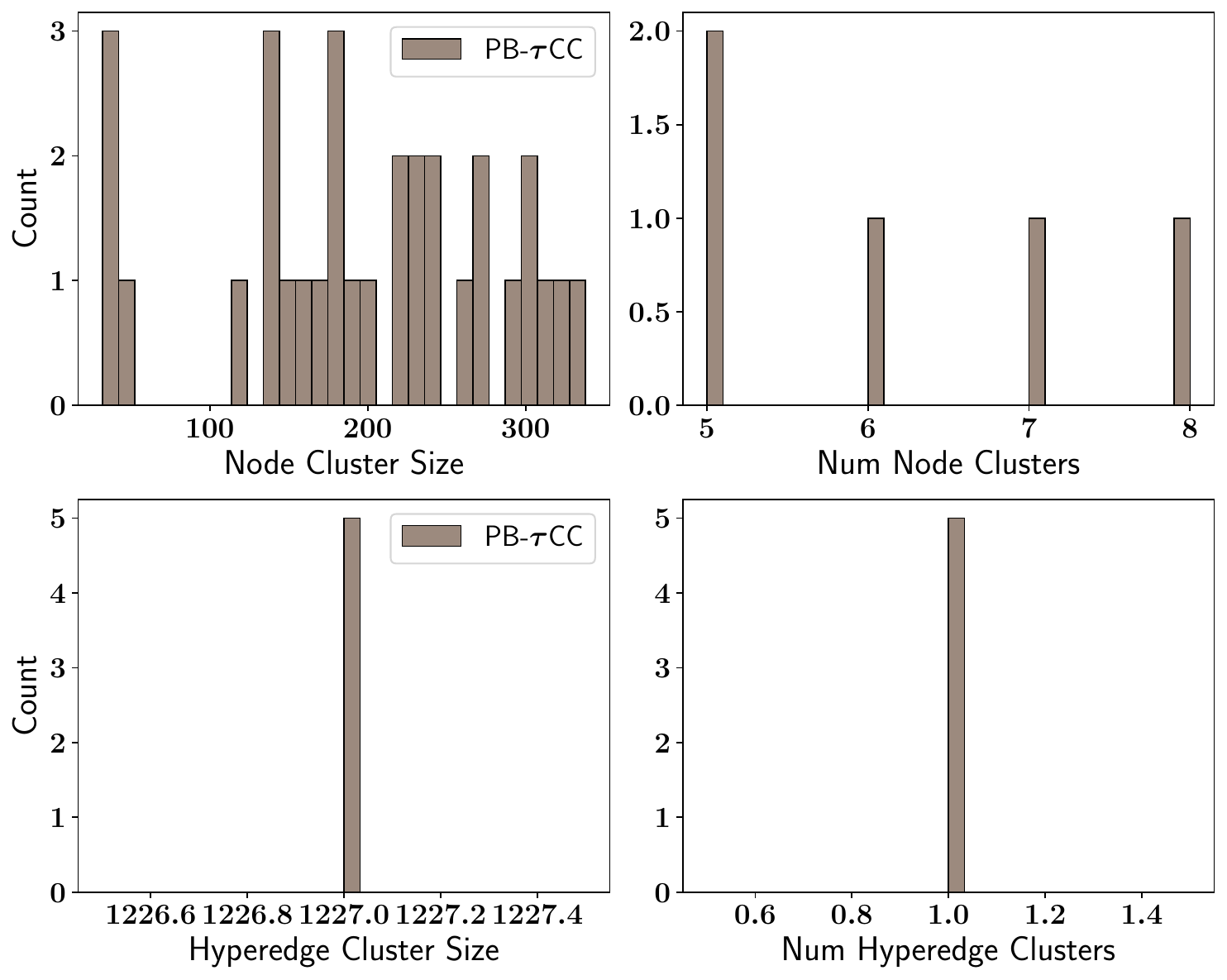}
        \caption{\cc}
    \end{subfigure}
    \begin{subfigure}{.32\linewidth}
    \centering
        \includegraphics[width=\linewidth]{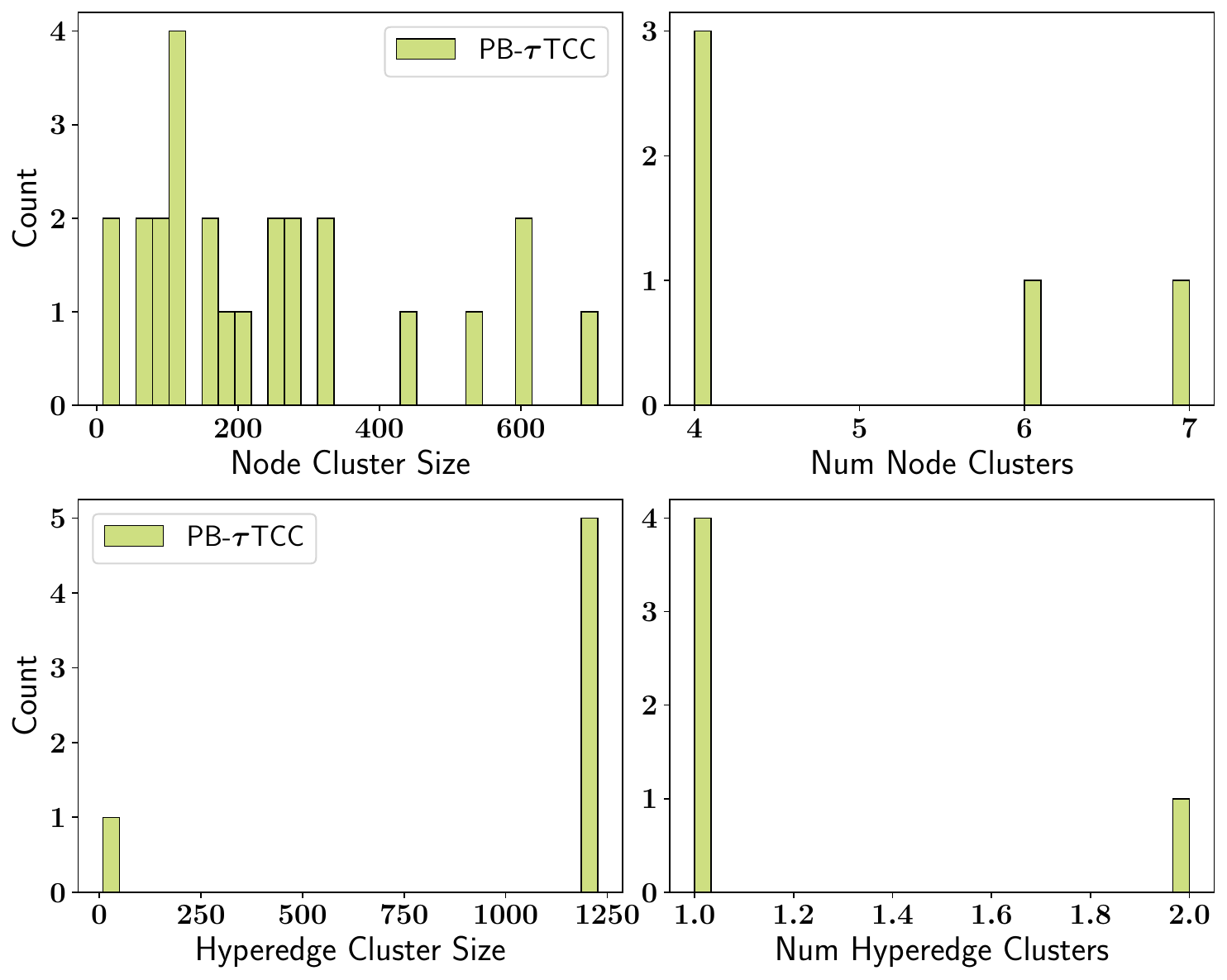}
        \caption{\tcc}
    \end{subfigure}
    \begin{subfigure}{.32\linewidth}
    \centering
        \includegraphics[width=\linewidth]{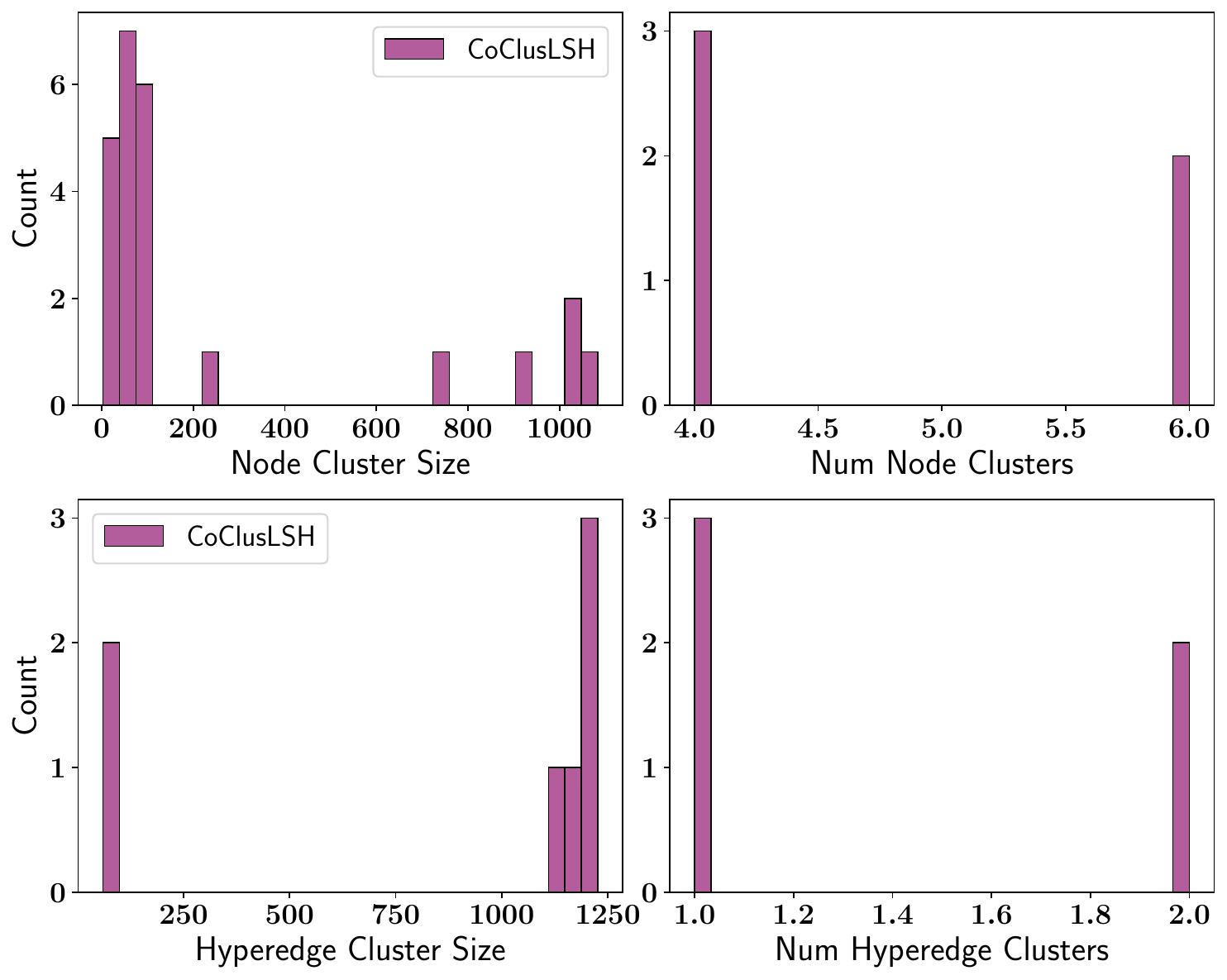}
        \caption{\leman}
    \end{subfigure}
    \begin{subfigure}{.32\linewidth}
    \centering
        \includegraphics[width=\linewidth]{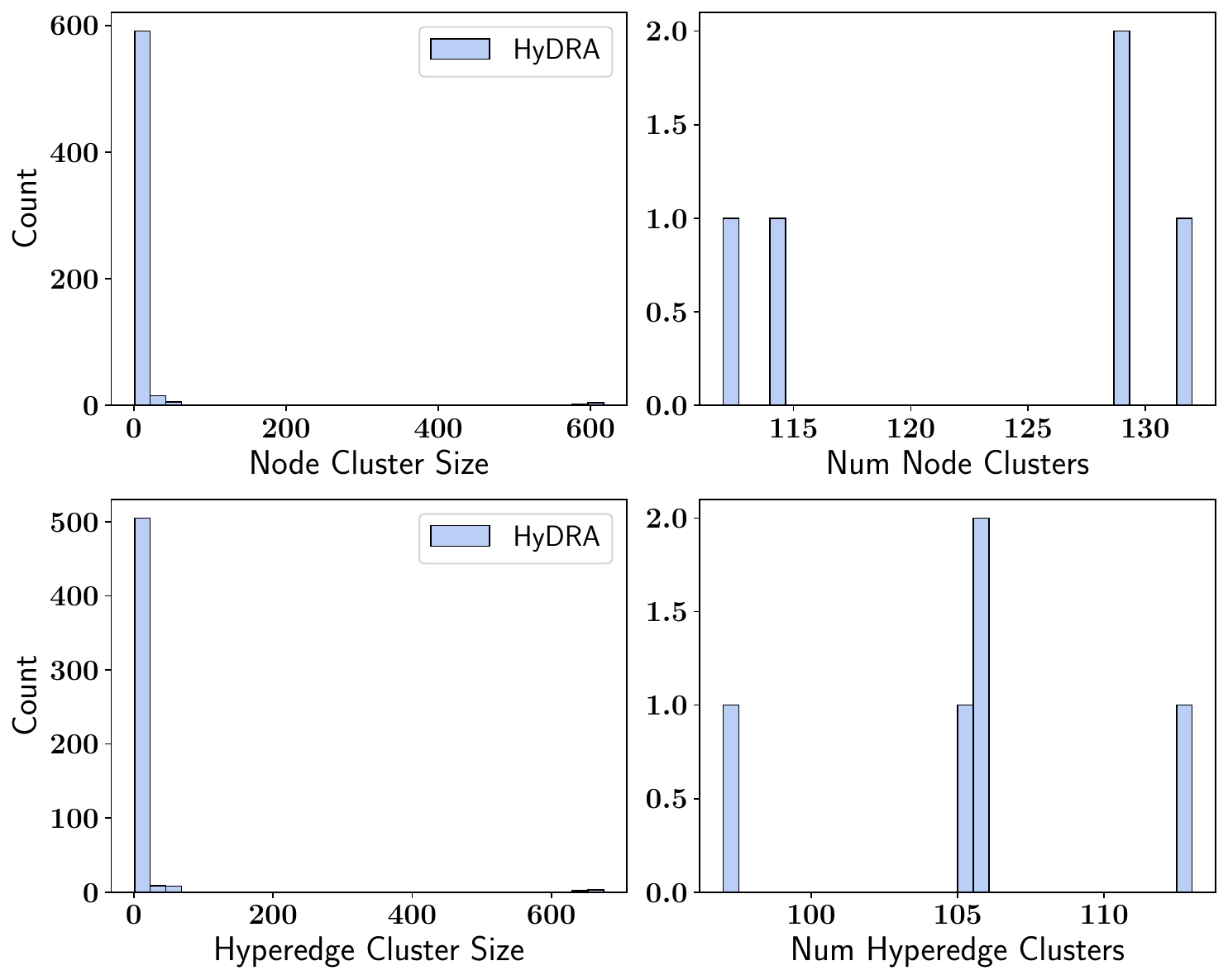}
        \caption{\our}
    \end{subfigure}
    \caption{Node and hyperedge cluster size distributions for summaries of \textsc{dblp}.}
    \label{fig:clusters_dblp}
\end{figure*}

Figures \ref{fig:clusters_senate} and \ref{fig:clusters_dblp} show the distributions of node and hyperedge cluster sizes (right plots), as well as the number of node and hyperedge clusters (left plots), for the summaries found by the different algorithms across two datasets.
Clear patterns emerge regarding how the algorithms summarize the hypergraphs.

\textbf{Prototype-based co-clustering and spectral \- clustering} methods tend to find very few clusters. \cc and \tcc often produce only 1–2 node and hyperedge clusters, sometimes with highly uneven sizes. 
SC and SB typically find a few more clusters, with most of them containing only a small number of nodes and hyperedges.
Hyperedge cluster sizes can still vary, but are often less extreme than in \cc and \tcc.

\textbf{\leman and \our} consistently find more clusters than the other algorithms. 
Both tend to produce cluster size distributions following a power law, with many small clusters and a few large ones. This behavior allows them to preserve finer details of the original hypergraph. Compared to \leman, \our usually identifies the largest number of node and hyperedge clusters, reflecting a stronger retention of the input’s structural patterns.

\spara{Summary.}
When the input hypergraph lacks a clear block structure, \our produces summaries that resemble the original data, preserving its structure while still achieving storage savings. In datasets with clearer structures, it still finds more clusters than the competitors, capturing more of the connectivity patterns. In contrast, \cc and \tcc tend to collapse nodes and hyperedges into very few clusters, often losing critical structural information.

\subsection{Queries Answering}
A key motivation for summarization is to accelerate query processing
while preserving accuracy.
As our summaries are lossless, one can always reconstruct the
original hypergraph and obtain exact query results.
In the following experiments, we are interested in the tradeoff between
speed and accuracy.
We therefore compare the results obtained by answering queries
\emph{approximately} and directly on the summary $(\supernodes, \superedges, \omega_\summary)$ without
reconstruction, against the exact results from the original hypergraph.
For clarity of presentation, we report results only for the five
smallest datasets.

\begin{figure}[t!]
    \centering
    \includegraphics[width=.9\linewidth]{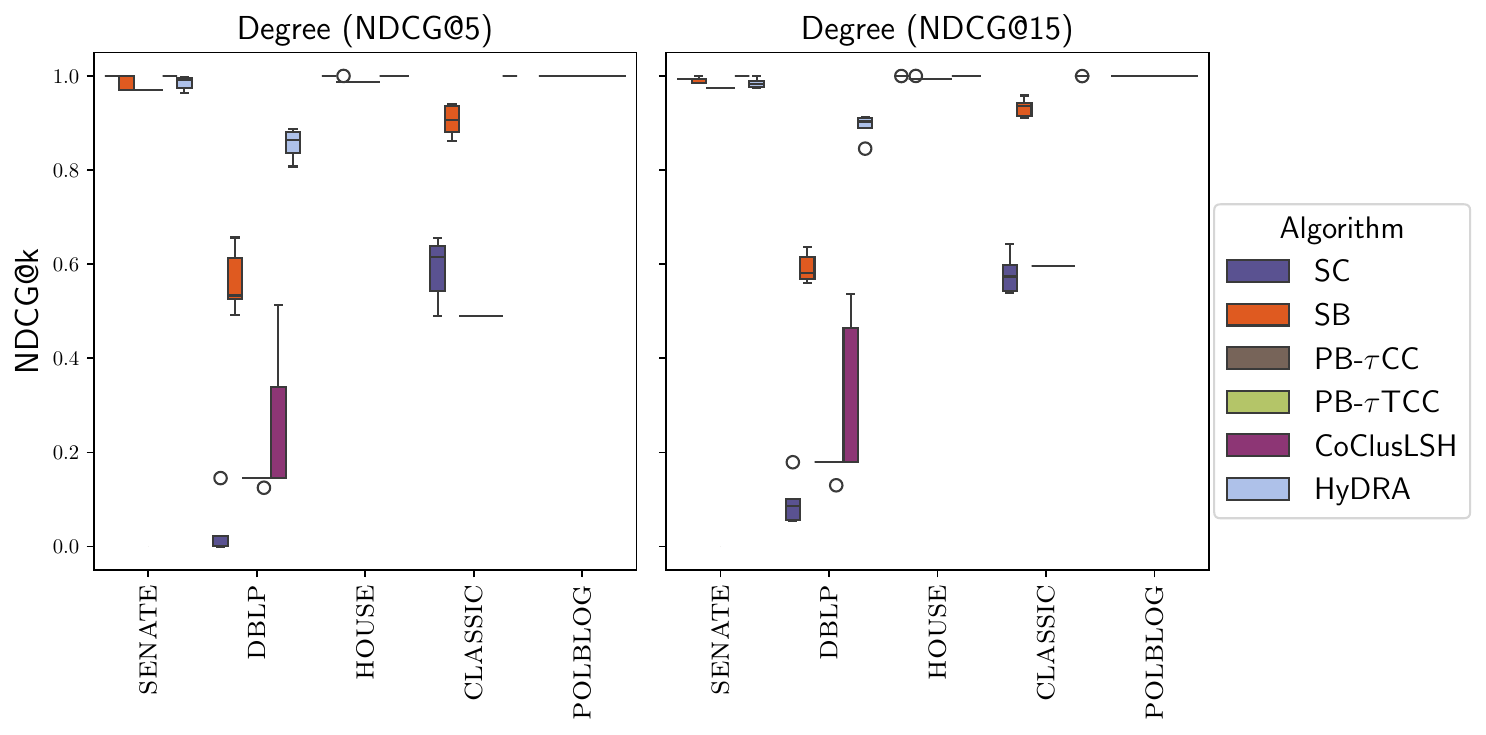}
    \includegraphics[width=.9\linewidth]{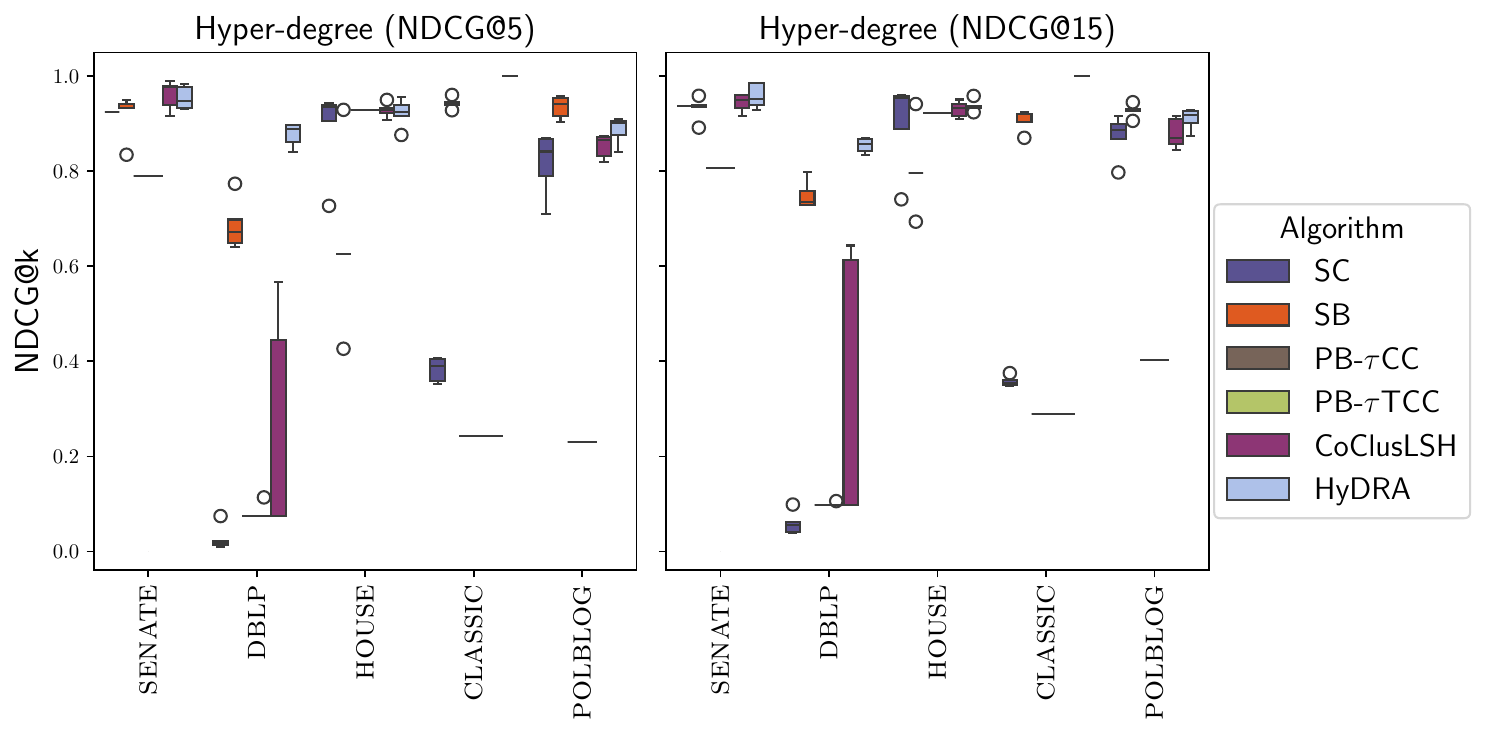}
    \caption{NDCG@k for node degree and hyper-degree. \label{fig:degree_q}}

    \vspace{2mm}

    \includegraphics[width=.9\linewidth]{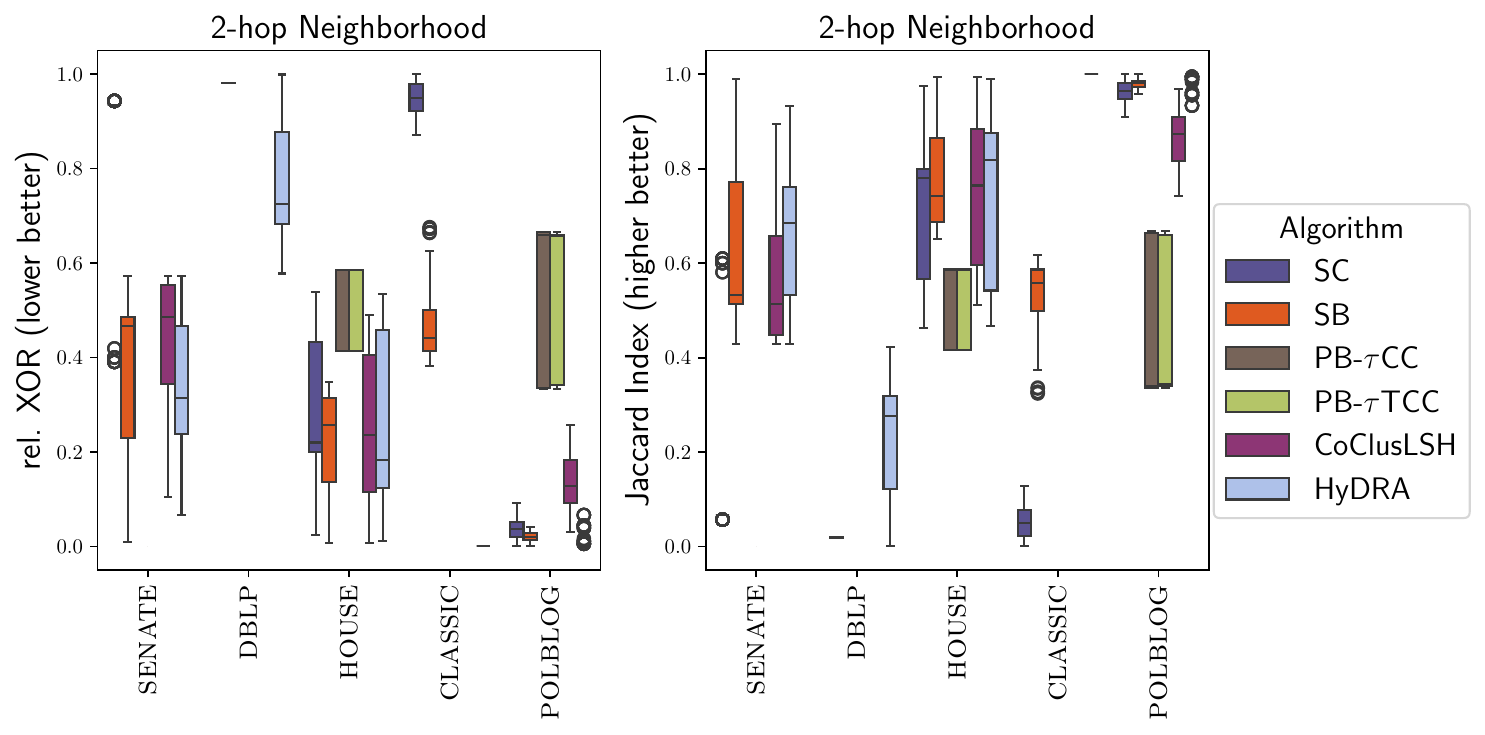}
    \caption{Relative XOR (left) and Jaccard Index (right) for $2$-hop reachability queries.   \label{fig:reachk_q}}
\end{figure}

\spara{Degree and hyper-degree.}
Let $\superedges_{\restriction v}$ denote the set of superhyperedges
including the supernode containing $v$.
Then, the hyper-degree of $v$ can be approximated as
$\hdegr{v} \approx \sum_{r \in \superedges_{\restriction v}}\omega(r)$,
and its degree can be approximated as
$\degr{v} \approx  \left|\bigcup_{r \in \superedges_{\restriction v}} \bigcup_{P \in r}P \right| - 1$.

We measure the quality of query answering using NDCG@k, with $k=5$ and
$k=15$.
This metric evaluates the alignment between the ranking of the top-$k$
nodes (by degree or hyper-degree) in the original hypergraph and in the
summary. Higher values indicate better alignment, especially in the
highest-ranked positions.

Our experiments show that \leman variants consistently outperform the
baselines.
In particular, our method achieves the best performance on most
datasets, with a significant advantage on \textsc{dblp}.
The exceptions are \textsc{classic}, where most algorithms do not yield
high-quality summaries due to the lack of a clear checkerboard
structure, and \textsc{polblog}, where SB tends to perform the best.

\begin{table}[t!]
   \footnotesize
    \caption{Mean and STD (in parentheses) of MAE and MSE across $100$ distance queries, and speedup factor (ratio between runtime for original hypergraph and for summary).}
    \label{fig:reach_mae_q}
   \resizebox{\linewidth}{!}{
    \begin{tabular}{l|lrcc|}
    \multicolumn{1}{c}{} & \multicolumn{1}{c}{\textbf{Algorithm}} & \textbf{Speedup} & \textbf{MAE} & \multicolumn{1}{c}{\textbf{MSE}} \\
    \cline{2-5}
    \multirow{6}{*}{\rotatebox[origin=c]{90}{\textsc{senate}}}
       & SC           & 2843.492 & \textbf{0.245} (0.037) & \textbf{0.245} (0.037) \\
       & SB           & 2684.542 & 0.871 (0.006) & 0.871 (0.006) \\
       & \cc          & 2304.719 & 0.996 (0.000) & 1.039 (0.000) \\
       & \tcc         & 2461.856 & 0.996 (0.000) & 1.039 (0.000) \\
       & \leman       & 2394.700 & 0.343 (0.041) & 0.347 (0.040) \\
       & \textbf{\our}         & 2004.613 & 0.314 (0.085) & 0.318 (0.086) \\
    \cline{2-5}
    \multirow{6}{*}{\rotatebox[origin=c]{90}{\textsc{dblp}}}
       & SC           & 701.966 & 2.088 (0.000) & 4.778 (0.000) \\
       & SB           & 346.577 & 2.098 (0.113) & 4.873 (0.492) \\
       & \cc           & 492.200 & 1.895 (0.065) & 4.116 (0.279) \\
       & \tcc          & 273.128 & 1.895 (0.105) & 4.049 (0.429) \\
       & \leman     & 642.080 & 2.116 (0.034) & 4.871 (0.190) \\
       & \textbf{\our}	       & 141.237 & \textbf{1.601} (0.125) & \textbf{3.919} (0.309) \\
    \cline{2-5}
    \multirow{6}{*}{\rotatebox[origin=c]{90}{\textsc{house}}}
        & SC          & 83687.437 & 0.282 (0.047) & 0.282 (0.047) \\
        & SB          & 70754.901 & \textbf{0.190} (0.008) & \textbf{0.190} (0.008) \\
        & \cc          & 86624.507 & 0.508 (0.000) & 0.508 (0.000) \\
        & \tcc         & 88145.201 & 0.508 (0.000) & 0.508 (0.000) \\
        & \leman     & 58183.153 & 0.221 (0.028) & 0.221 (0.028) \\
        & \textbf{\our}      & 64830.843 & 0.248 (0.049) & 0.248 (0.049) \\
    \cline{2-5}
    \multirow{6}{*}{\rotatebox[origin=c]{90}{\textsc{classic}}}
        & SC         & 615.304 & 1.235 (0.180) & 2.064 (0.645) \\
        & SB         & 21.908 & 0.569 (0.048) & 0.629 (0.070) \\
        & \cc         & 2981.983 & 1.260 (0.059) & 1.780 (0.177) \\
        & \tcc        & 2891.427 & 1.221 (0.045) & 1.663 (0.136) \\
        & \leman	  & 3234.043 & 1.411 (0.003) & 2.234 (0.009) \\
        & \textbf{\our}        & 1.060 & \textbf{0.001} (0.002) & \textbf{0.001} (0.002) \\
    \cline{2-5}
    \multirow{6}{*}{\rotatebox[origin=c]{90}{\textsc{polblog}}}
        & SC          & 99318.237 & 0.027 (0.007) & 0.027 (0.007) \\
        & SB          & 44061.004 & 0.015 (0.003) & 0.015 (0.003) \\
        & \cc         & 352246.976 & 0.539 (0.004) & 0.539 (0.004) \\
        & \tcc        & 342754.047 & 0.534 (0.004) & 0.534 (0.004) \\
        & \leman	  & 319933.815  & 0.103 (0.020) & 0.103 (0.020) \\
        & \textbf{\our}        & 36.293 & \textbf{0.008} (0.005) & \textbf{0.008} (0.005) \\
    \cline{2-5}
    \end{tabular}
    }
    \end{table}

\spara{Distance and reachability.}
We next consider queries related to node connectivity. The
\emph{distance} query asks for the shortest path between two nodes,
and the \emph{reachability} query asks whether one node can reach
another. Given a pair of nodes $u,v \in \nodes$, we first identify the
corresponding supernodes $P_u, P_v \in \supernodes$ such that
$u \in P_u$ and $v \in P_v$. We then run a shortest-path (or
reachability) algorithm on the summary using $P_u$ and $P_v$ as inputs.
To evaluate performance, we randomly sampled $100$ nodes (for $2$-hop
neighborhood comparisons) and $100$ node pairs (for distance and reachability). 
The outputs are evaluated in terms of (1) reachability errors,
that is, percentage of incorrect classifications of node pairs as
connected/disconnected; (2) MAE and MSE of the shortest-path
distances; and (3) Jaccard similarity and relative XOR between
the sets of 2-hop neighbors in the original hypergraph and the summary.

For reachability, \leman and \our incur fewer errors
than all baselines, with \our achieving the lowest error rate overall.
Specifically, $0.299$ for SC, $0.274$ for SB, $0.312$ for \cc, $0.329$ for \tcc, $0.165$ for \leman, and $0.144$ for \our.
For distance queries (\Cref{fig:reach_mae_q}), our method delivers the
best performance on three out of five datasets, with spectral clustering
algorithms (SC, SB) performing better on the remaining two.
For neighborhood overlap (\Cref{fig:reachk_q}), our method consistently
achieves the highest Jaccard similarity and lowest XOR difference.

Overall, even though  \cc and \tcc are faster, \our achieves the best balance between accuracy and efficiency, answering connectivity queries with high accuracy.

\begin{figure}[t!]
    \centering
    \begin{tabular}{cc}
        \includegraphics[width=0.46\linewidth]{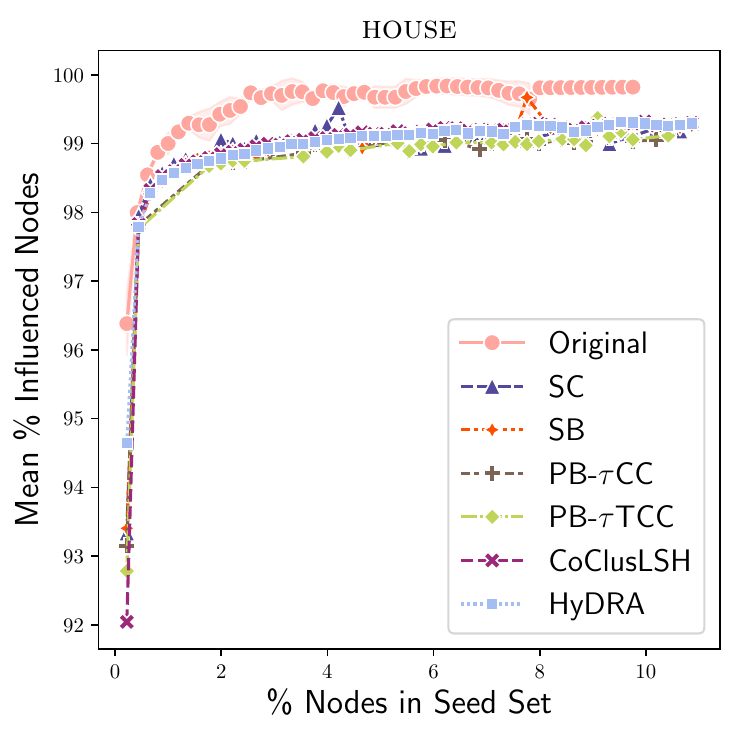} &
        \includegraphics[width=0.46\linewidth]{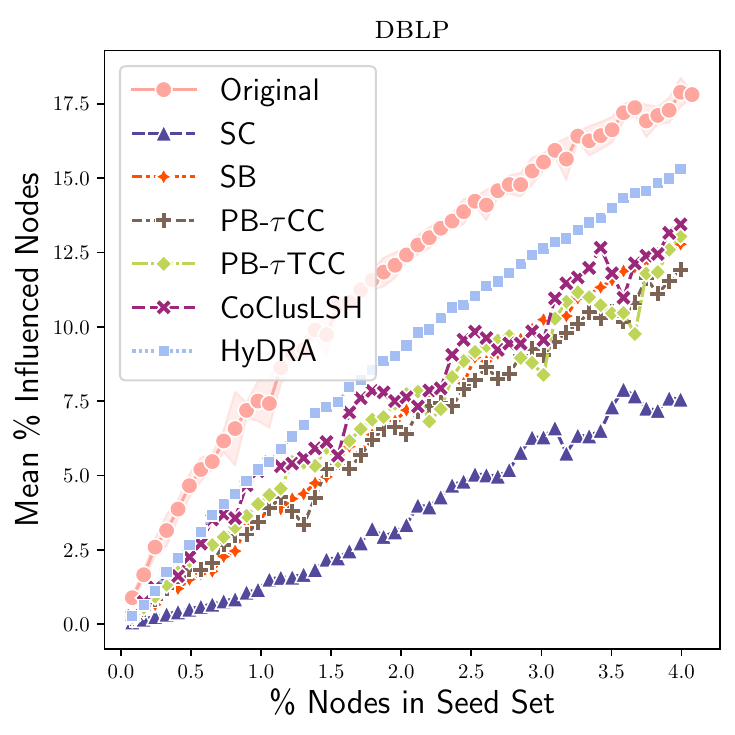}
    \end{tabular}   
    \begin{tabular}{cc}
        \includegraphics[width=0.46\linewidth]{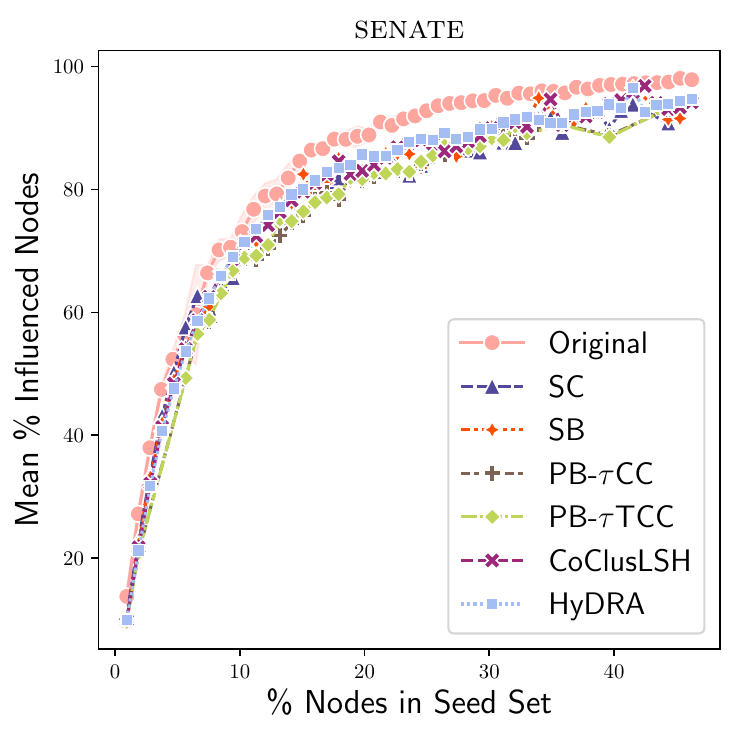} &
        \includegraphics[width=0.46\linewidth]{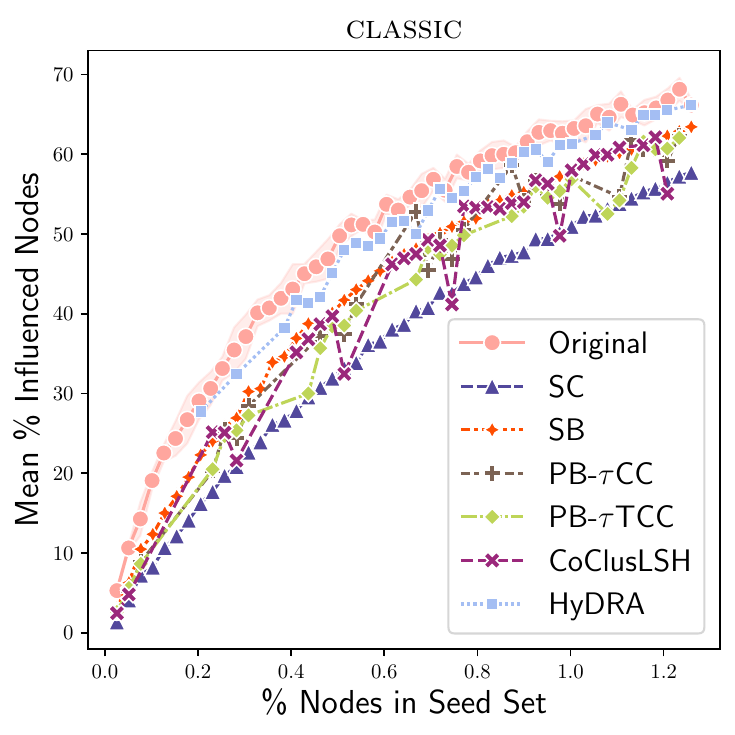}
    \end{tabular}
    \caption{Mean percentage of influenced nodes for varying percentage of nodes in the seed set, for the original hypergraphs and the corresponding summaries.}
    \label{fig:im_results}
\end{figure}

\mpara{Connectivity.} 
A connected component represents an isolated group of nodes that can
reach each other but not nodes outside the component. 
In many domains (e.g., social, biological, or communication networks),
connected components correspond to real-world clusters or communities.
Preserving them is important: merging distinct components may falsely
suggest that the system is more cohesive or resilient than it truly is.  

\begin{figure}[t]
    \centering
    \includegraphics[width=\linewidth]{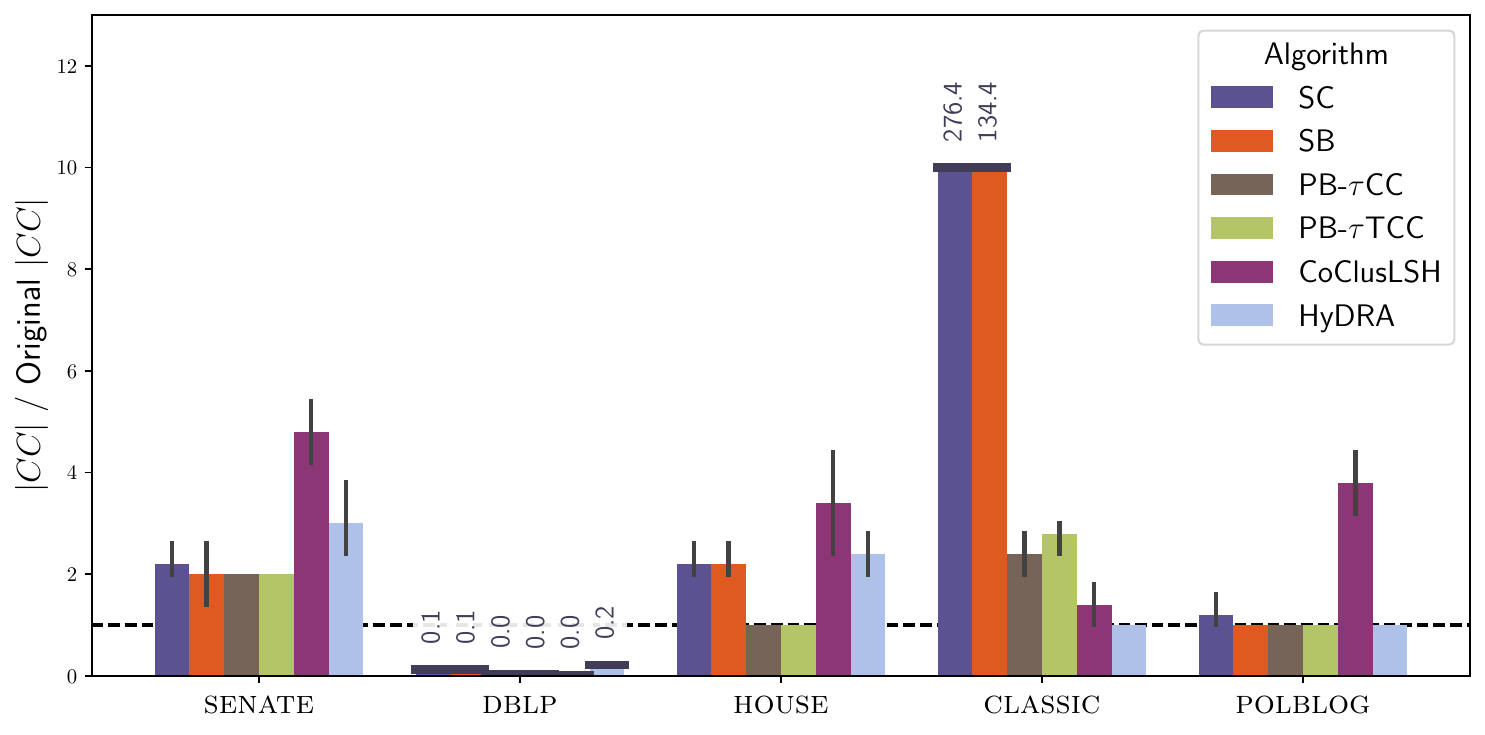}
    \caption{Ratio between the number of connected components of the summaries and that of the original hypergraphs.}
    \label{fig:cc_q}
\end{figure}

\Cref{fig:cc_q} shows the ratio between the number of connected
components in the summary and in the original hypergraph. 
Ratios below 1 indicate that the summary has fewer components than the
original hypergraph, while ratios greater than 1 indicate that the
summary is more disconnected than the original. 
The results highlight different behaviors across algorithms. Spectral
methods (SC, SB) tend to fragment the hypergraph, producing summaries
with more connected components than the original. \cc and \tcc typically
yield the opposite effect, producing fewer but larger components due to
their coarser clustering of nodes into large supernodes. 
Our approach consistently preserves connectivity patterns better than
its original variant \leman, striking a balance between avoiding
spurious connections and limiting artificial fragmentation.

\mpara{Closeness centrality.}
The closeness centrality of a node $v$ is defined as
$\frac{n-1}{\sum_{u \in \nodes} \text{dist}(v,u)}$ if $v$ is connected, and $0$ otherwise.

To approximate the centrality of a node in the summary, we compute the
centrality of the supernode $P_v \in \supernodes$ containing it. 
Then, we evaluate ranking quality using NDCG@5 and NDCG@15.
The results are reported in \Cref{tab:closeness}.

As summaries may disconnect parts of the hypergraph, centrality scores
can drop to $0$ for nodes that are disconnected in the summary. 
In our experiments, we measured the percentage of query nodes that
became disconnected in the summaries:  
\leman: $44.6\%$, \our: $12.1\%$, \cc: $56.8\%$, \tcc: $56.8\%$, SC: $44.5\%$, and SB: $30.8\%$. 
As we can see, \our substantially reduces the fraction of nodes that
become disconnected in the summary.
This better connectivity preservation generally leads to improvements in
centrality ranking quality on larger datasets: \our outperforms other
methods on the larger instances, although the gains are small.
On smaller datasets, \cc and some spectral methods occasionally match or
slightly exceed \leman and \our.

However, because \our typically produces a larger number of clusters,
its summaries contain a number of hyperedges comparable to the original
hypergraph, resulting in smaller time savings when running centrality
queries. In contrast, other methods (\cc, \tcc, SC, SB) save more time
but at the cost of losing accuracy because of the disconnected nodes.

In short, \our trades reduced time savings for better preservation of
connectivity and slightly better ranking accuracy on larger datasets;
other methods save more time but generally at the expense of
disconnected nodes and lower accuracy.

\begin{table}[t]
\centering
\caption{Closeness centrality queries: ranking quality (NDCG) and speedup factor (ratio between runtime for original hypergraph and for summary).}
\label{tab:closeness}
\resizebox{\linewidth}{!}{
    \begin{tabular}{l|lrcc|}
    \multicolumn{1}{c}{} & \textbf{Algorithm} & \textbf{Speedup} & \textbf{NDCG@5} & \multicolumn{1}{c}{\textbf{NDCG@15}} \\
    \cline{2-5}
    \multirow[c]{6}{*}{\rotatebox[origin=c]{90}{\textsc{senate}}} 
     & SC & 6281.168 & 0.971 & 0.977 \\
     & SB & 3759.239 & 0.974 & 0.975 \\
     & \cc & 6349.201 & \textbf{0.985} & \textbf{0.985} \\
     & \tcc & 6984.313 & \textbf{0.985} & \textbf{0.985} \\
     & \leman & 2664.034 & 0.977 & 0.980 \\
     & \textbf{\our} & 711.275 & 0.978 & 0.984 \\
    \cline{2-5}
    \multirow[c]{6}{*}{\rotatebox[origin=c]{90}{\textsc{dblp}}} 
     & SC & 3550112.853 & \textbf{0.001} & 0.001 \\
     & SB & 1067882.490 & \textbf{0.001} & 0.001 \\
     & \cc & 4193790.172 & \textbf{0.001} & \textbf{0.002} \\
     & \tcc & 609097.663 & \textbf{0.001} & \textbf{0.002} \\
     & \leman & 1607992.342 & \textbf{0.001} & 0.001 \\
     & \textbf{\our} & 34748.996 & \textbf{0.001} & \textbf{0.002} \\
    \cline{2-5}
    \multirow[c]{6}{*}{\rotatebox[origin=c]{90}{\textsc{house}}} 
     & SC & 87679.457 & \textbf{1.000} & \textbf{1.000} \\
     & SB & 87493.922 & 0.999 & 0.999 \\
     & \cc & 141191.818 & 0.998 & 0.999 \\
     & \tcc & 149983.537 & 0.998 & 0.999 \\
     & \leman & 3164.949 & 0.999 & \textbf{1.000} \\
     & \textbf{\our} & 5012.706 & 0.999 & \textbf{1.000} \\
    \cline{2-5}
    \multirow[c]{6}{*}{\rotatebox[origin=c]{90}{\textsc{classic}}} 
     & SC & 10037.308 & 0.800 & 0.825 \\
     & SB & 6.441 & 0.939 & 0.934 \\
     & \cc & 1673690.339 & 0.798 & 0.829 \\
     & \tcc & 1347895.609 & 0.798 & 0.829 \\
     & \leman & 1191444.102 & 0.773 & 0.801 \\
     & \textbf{\our} & 1.139 & \textbf{1.000} & \textbf{1.000} \\
    \cline{2-5}
    \multirow[c]{6}{*}{\rotatebox[origin=c]{90}{\textsc{polblog}}} 
     & SC & 1158.150 & \textbf{1.000} & \textbf{1.000} \\
     & SB & 81.734 & \textbf{1.000} & \textbf{1.000} \\
     & \cc & 384232.517 & \textbf{1.000} & \textbf{1.000} \\
     & \tcc & 460207.653 & \textbf{1.000} & \textbf{1.000} \\
     & \leman & 391161.280 & \textbf{1.000} & \textbf{1.000} \\
     & \textbf{\our} & 1.076 & \textbf{1.000} & \textbf{1.000} \\
    \cline{2-5}
    \end{tabular}
}
\end{table}

\mpara{Influence maximization (IM).}
IM seeks a set of nodes in a network
that, when initially activated, spreads the most influence.
It has been studied extensively in graphs; yet, hypergraphs allow for more realistic modeling of higher-order contagion processes.

Genetti et al.~\cite{genetti2024influence} formulate hypergraph IM as a bi-objective optimization problem that jointly maximizes influence and minimizes seed set size. 
Their algorithm, HN-MOEA, uses evolutionary optimization with hypergraph-aware operators to efficiently explore the solution
space and return a Pareto front of solutions with various tradeoffs between influence spread and seed set size.
Influence spread is evaluated using the Susceptible–Infected model with Contact Process dynamics (SICP)~\cite{xie2022influence}, where infected nodes repeatedly select an incident hyperedge and infect all susceptible nodes in it with probability~$\beta$.

We run HN-MOEA on both original hypergraphs and
their summaries, with a maximum seed set size of $50$. For each
solution on a summary, seed sets are instantiated from the corresponding supernodes and evaluated under SICP using $100$ Monte Carlo trials with infection probabilities
$\beta \in \{0.005, 0.02\}$.

\Cref{fig:im_results} reports the mean percentage of
influenced nodes at the stationarity for different seed set sizes
(reported as percentages with respect to the number of nodes in the
hypergraph). Results show that summaries produced by \our yield diffusion dynamics
closest to the original, particularly in datasets with stronger
checkerboard structures (e.g., \textsc{house}).
For \textsc{dblp}, the percentage of influenced
nodes is lower than in the original hypergraphs; however, the
competitors deviate even more, making the advantage of our approach more
pronounced in these cases.

\begin{figure}[t!]
    \centering
    \includegraphics[width=.85\linewidth]{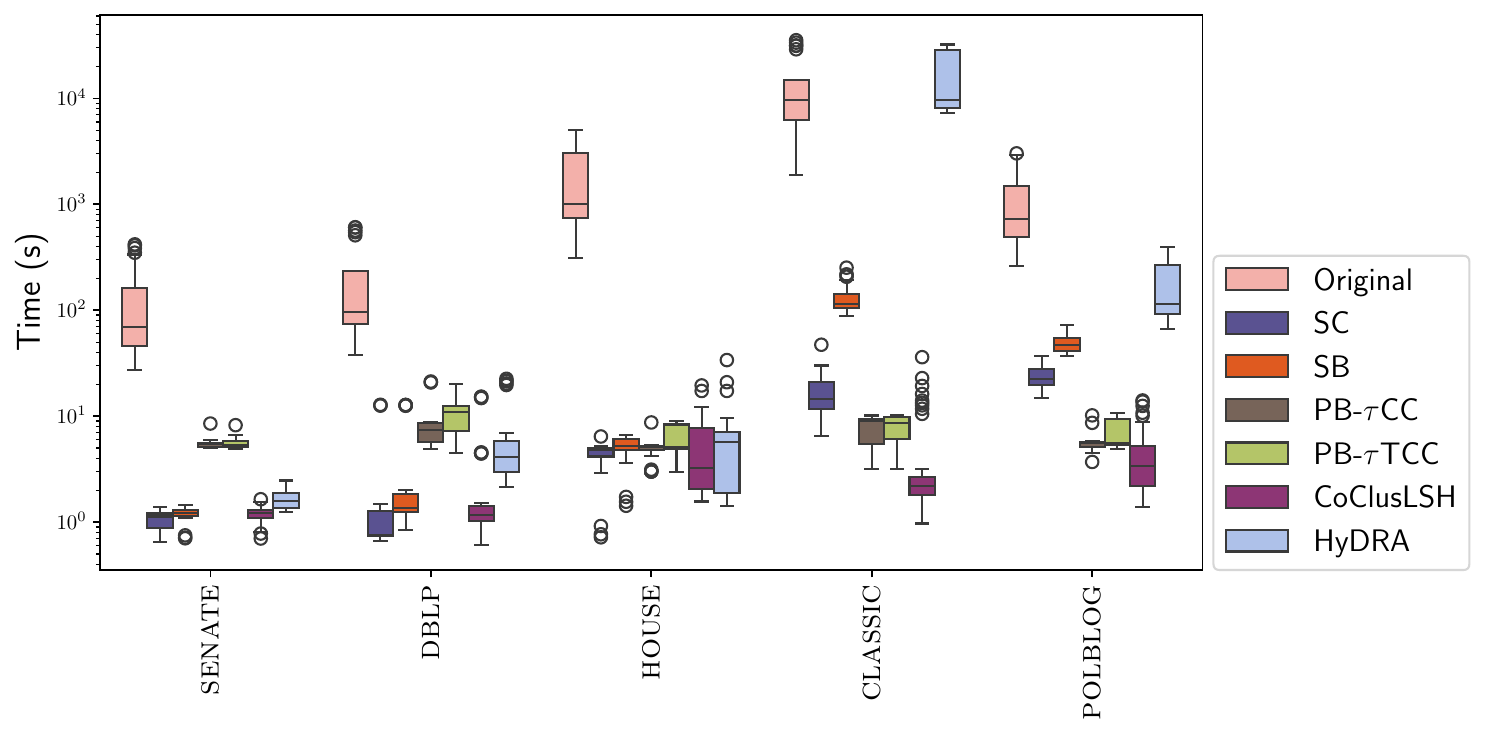}
    \vspace{-4mm}
    \caption{Running times (log scale) of the IM processes in the original hypergraphs and the corresponding summaries.}
    \label{fig:im_runtimes}
    \vspace{-4mm}
\end{figure}

Runtimes for HN-MOEA on summaries (\Cref{fig:im_runtimes}) are significantly reduced 
compared to the original hypergraphs, except for \our on
\textsc{classic}, where limited compressibility yields summaries close in size to the original.

\subsection{Hyperparameter Analysis}
\label{sec:app-hyperparameters}

\begin{table*}[!ht]
\centering
\footnotesize
\caption{\our: Hyper-parameter analysis.}
\label{tbl:hyp-analysis}
    \begin{tabular}{llrrr c lrrrr}
    \toprule
    & \textbf{Parameter} & \textbf{Value} & \textbf{Time (s)} & \textbf{Cost Imp.} & & \textbf{Parameter} & \textbf{Value} & \textbf{Time (s)} & \textbf{Cost Imp.} \\
    \midrule
    \multirow{10}{*}{\rotatebox{90}{\textsc{senate}}} & \multirow{3}{*}{Signature Size} & 5 & 6.179 & 0.8175 & & \multirow{3}{*}{Signature Size} & 5 & 72.369 & 0.5642 & \multirow{10}{*}{\rotatebox{90}{\textsc{dblp}}} \\
    & & 10 & 7.079 & 0.8172 & & & 10 & 148.850 & 0.6037 & \\
    & & 20 & 12.161 & 0.8156 & & & 20 & 201.087 & 0.6144 & \\
    \cline{3-5}\cline{8-10}
    & \multirow{3}{*}{Hash Tables} & 5 & 12.161 & 0.8156 & & \multirow{3}{*}{Hash Tables} & 5 & 201.087 & 0.6144 & \\
    & & 10 & 8.774 & 0.8073 & & & 10 & 241.290 & 0.6145 & \\
    & & 20 & 13.880 & 0.8105 & & & 20 & 345.057 & 0.6162 & \\
    \cline{3-5}\cline{8-10}
    & \multirow{4}{*}{Cost Thresh.} & 1e-4 & 7.516 & 0.8156 & & \multirow{4}{*}{Cost Thresh.} & 1e-4 & 192.494 & 0.6144 & \\
    & & 1e-2 & 7.557 & 0.8144 & & & 1e-2 & 157.392 & 0.6022 & \\
    & & 1 & 12.161 & 0.8156 & & & 1 & 201.087 & 0.6144 & \\
    \midrule
    \multirow{10}{*}{\rotatebox{90}{\textsc{house}}} & \multirow{3}{*}{Signature Size} & 5 & 23.734 & 0.8577 & & \multirow{3}{*}{Signature Size} & 5 & 424.736 & 0.0004 & \multirow{10}{*}{\rotatebox{90}{\textsc{classic}}}\\
    & & 10 & 43.872 & 0.8668 & & & 10 & 1336.222 & 0.0009 \\
    & & 20 & 31.841 & 0.8654 & & & 20 & 1221.691 & 0.0035 \\
    \cline{3-5}\cline{8-10}
    & \multirow{3}{*}{Hash Tables} & 5 & 31.841 & 0.8654 & & \multirow{3}{*}{Hash Tables} & 5 & 1221.691 & 0.0035 \\
    & & 10 & 53.616 & 0.8711 & & & 10 & 1681.726 & 0.0026 \\
    & & 20 & 62.713 & 0.8695 & & & 20 & 2872.132 & 0.0103 \\
    \cline{3-5}\cline{8-10}
    & \multirow{4}{*}{Cost Thresh.} & 1e-4 & 27.608 & 0.8654 & & \multirow{4}{*}{Cost Thresh.} & 1e-4 & 693.616 & 0.0024 \\
    & & 1e-2 & 22.409 & 0.8645 & & & 1e-2 & 193.074 & 0.0000 \\
    & & 1 & 31.841 & 0.8654 & & & 1 & 1221.691 & 0.0035 \\
    \midrule
    \multirow{10}{*}{\rotatebox{90}{\textsc{polblog}}} & \multirow{3}{*}{Signature Size} & 5 & 378.643 & 0.3715 & & & & & \\
    & & 10 & 526.014 & 0.3697 & & & & & \\
    & & 20 & 736.129 & 0.3705 & & & & & \\
    \cline{3-5}
    & \multirow{3}{*}{Hash Tables} & 5 & 736.129 & 0.3705 & & & & & \\
    & & 10 & 1011.073 & 0.3642 & & & & & \\
    & & 20 & 1056.550 & 0.3738 & & & & & \\
    \cline{3-5}
    & \multirow{4}{*}{Cost Thresh.} & 1e-4 & 614.539 & 0.3700 & & & & & \\
    & & 1e-2 & 325.719 & 0.3605 & & & & & \\
    & & 1 & 736.129 & 0.3705 & & & & & \\
    \bottomrule
    \end{tabular}
\end{table*}

Our algorithm involves several hyperparameters that affect the tradeoff
between summary quality and runtime. This analysis focuses on the
parameters related to the hashing procedure used to generate candidate
cluster pairs and the stopping criterion that governs the merging
process. 
We tested different parameter settings on a subset of datasets.
Results are reported in \Cref{tbl:hyp-analysis}.

\mpara{Hashing parameters.} 
\Cref{alg:mergeclusters} relies on hashing to identify candidate merges. 
Two parameters control this step: (1) the \emph{signature size} $r$,
which regulates how much information is encoded in each hash, and
(2) the \emph{number of hash tables} $b$, which determines how many
independent hashings are used in parallel. 
Larger signature sizes reduce the chance that two clusters hash to the
same bucket, which leads to a greater number of candidate groups to
process. This, in turn, can result in fewer merges per iteration, as
there are fewer overlapping candidates. Experimentally, we observed that
increasing the signature size tends to slightly increase runtime but
also leads to modest improvements in the final cost.  

\mpara{Stopping criterion.} 
The merging process in \Cref{alg:algorithm} is controlled by the \emph{cost threshold} $\tau$. 
After each row and column merge attempt, we compute the relative cost improvement. 
If the improvement falls below the threshold, the merge attempt is
considered unsuccessful. If both row and column merge attempts fail, the
procedure terminates.  
We consider two types of thresholds: (i) an absolute threshold of $1$,
which requires an improvement of at least one unit of cost, and (ii)
relative thresholds $<1$, where the improvement is measured as
$(c_{t-1} - c_t) / c_{t-1}$. 
Smaller thresholds allow more iterations and can lead to better cost
improvements, but at the expense of higher runtime. Conversely, larger
thresholds save time but may stop the process prematurely, sometimes
missing beneficial merges.

\noindent 
Although the runtime varied substantially with the choice of parameters,
the cost improvements did not show consistent trends across datasets.
In the following experiments, we fixed $r=20$ and $b=5$ for both \leman
and \our, and $\tau = 1$.

\subsection{Discussion}\label{sec:discussion}
Our experiments show that co-clustering formulations can be adapted to the task of lossless hypergraph summarization. By optimizing an objective that directly targets storage efficiency while preserving losslessness, we bridge co-clustering and summarization, producing compact summaries that fully reconstruct the original hypergraph while substantially reducing storage costs.

Although our approach incurs higher computational cost than
prototype- or spectral-based competitors, this overhead yields substantially better summaries---both in terms of
structure and accuracy.
Random baselines, though extremely fast, often produce degenerate summaries that can exceed the size of the original
data (e.g., on \textsc{NDC-classes} and \textsc{DAWN}), underscoring the need for principled optimizations.
Spectral algorithms achieve competitive quality but require the number of node clusters as input, which limits their practicality and often leads to suboptimal hyperedge groupings, as they are not designed for hypergraph summarization.
Prototype-based approaches, though parameter-free and efficient, tend to collapse the structure into a few oversized clusters, losing much of the original connectivity.
\leman is much superior, managing to trade off running time for
accuracy. However, it tends to create large clusters (see \Cref{fig:clusters_senate} and \Cref{fig:clusters_dblp}), which can degrade performance on certain queries (e.g., connectivity, influence maximization).

In contrast, \our strikes a favorable balance between summary
compactness and fidelity. It generates finer-grained node and hyperedge clusters that better preserve structural diversity.  While this granularity reduces query-time speedups, it yields more accurate query answers, particularly for distance, degree, and reachability queries.
The influence maximization case study highlights an appealing
application of our summaries: they can be leveraged to identify
high-quality seed sets orders of magnitude faster than operating on the raw hypergraphs, confirming that lossless summaries can serve as effective surrogates for large-scale analysis. 

\section{Conclusions}\label{sec:conclusions}
We presented \our, the first framework for lossless hypergraph summarization.
By framing the task as a storage-aware co-clustering problem, we designed an efficient greedy algorithm that produces compact summaries capable of exactly reconstructing the original hypergraph.
Extensive experiments on diverse real-world hypergraphs demonstrate that \our outperforms random, spectral, and prototype-based baselines in both compactness and fidelity, achieving up to 93\% reduction in storage cost while preserving key structural and connectivity patterns.
These summaries enable accurate and efficient downstream analyses, including distance, degree, and reachability queries, and the influence maximization case study proves their practical utility for large-scale hypergraph analytics.

This work establishes a principled link between co-clustering and hypergraph summarization, opening new directions for scalable analysis over complex relational data.
Future work includes extending \our to dynamic or attributed hypergraphs and integrating the summaries into higher-order graph learning pipelines.

\bibliographystyle{spbasic}      
\bibliography{refs}

\end{document}